\documentclass[sigconf]{acmart}
\usepackage{booktabs} 
\usepackage{multirow} 
\usepackage{graphicx} 
\usepackage{tabularx}

\renewcommand\footnotetextcopyrightpermission[1]{}
\setcopyright{none}

\AtBeginDocument{%
  }
\usepackage{algorithm}
\usepackage{enumitem}
\usepackage{multirow}
\usepackage{amsmath}
\usepackage{algorithmic}
\usepackage{listings}
\usepackage{subfigure}
\usepackage{graphicx}
\usepackage{blindtext}
\usepackage{array}

\author{Shou Liu}
\email{liushuo22s@ict.ac.cn}
\affiliation{
    \institution{University of Chinese Academy of Sciences}
    \country{China}
}
\author{Di Yao$^\dagger$}
\email{yaodi@ict.ac.cn}
\affiliation{
    \institution{Institute of Computing Technology, Chinese Academy of Sciences}
    \country{China}
}
\author{Yan Lin}
\email{lyan@cs.aau.dk}
\affiliation{
    \institution{Department of Computer Science, Aalborg University}
    \country{Denmark}
}
\author{Gao Cong}
\email{gaocong@ntu.edu.sg}
\affiliation{
    \institution{College of Computing and Data Science, Nanyang Technological University}
    \country{Singapore}
}
\author{Jingping Bi}
\email{bjp@ict.ac.cn}
\affiliation{
    \institution{Institute of Computing Technology, Chinese Academy of Sciences}
    \country{China}
    }

\thanks{$^\dagger$ Corresponding authors.}

\newcommand{\ie}{\emph{i.e.}\xspace} 
\newcommand{\etc}{\emph{etc.}\xspace} 
\newcommand{\eg}{\emph{e.g.}\xspace} 
\newcommand{\bm}{\mathbf}

\newcommand{\our}{{\texttt{Traj-MLLM}}\xspace}

\begin{document}
\title{\our: Can Multimodal Large Language Models Reform Trajectory Data Mining?}

\begin{abstract}
Building a general model capable of analyzing human trajectories across different geographic regions and different tasks becomes an emergent yet important problem for various applications. However, existing works suffer from the generalization problem, \ie, they are either restricted to train for specific regions or only suitable for a few tasks. Given the recent advances of multimodal large language models (MLLMs), we raise the question: can MLLMs reform current trajectory data mining and solve the problem? Nevertheless, due to the modality gap of trajectory, how to generate task-independent multimodal trajectory representations and how to adapt flexibly to different tasks remain the foundational challenges. In this paper, we propose \our, which is the first general framework using MLLMs for trajectory data mining. By integrating multiview contexts, \our transforms raw trajectories into interleaved image-text sequences while preserving key spatial-temporal characteristics, and directly utilizes the reasoning ability of MLLMs for trajectory analysis. Additionally, a prompt optimization method is proposed to finalize data-invariant prompts for task adaptation. Extensive experiments on four publicly available datasets show that \our outperforms state-of-the-art baselines by $48.05\%$, $15.52\%$, $51.52\%$, $1.83\%$ on travel time estimation, mobility prediction, anomaly detection and transportation mode identification, respectively. \our achieves these superior performances without requiring any training data or fine-tuning the MLLM backbones.
\end{abstract}


\maketitle
\section{Introduction}
Technical advances in GPS positioning enable the collection of world-wide human trajectory data derived from mobile devices~\cite{zheng2011geolife}, ride-hailing services~\cite{sun2022optimizing}, \etc. 
Trajectory data mining has become an increasingly important research topic, providing tools for numerous tasks such as travel time estimation~\cite{jiang2023self,yang2023lightpath,zhou2024plm4traj}, trajectory prediction~\cite{feng2018deepmove, zhou2024plm4traj}, anomaly detection~\cite{liu2020online, han2022deeptea, wang2024multi}, transportation mode identification~\cite{jiang2017trajectorynet,dabiri2019semi,liang2022trajformer}, \etc. However, most works~\cite{feng2018deepmove, liu2020online, han2022deeptea, wang2024multi} are tailored to specific tasks and regions hindering their usage in real-world scenarios. Building a unified model capable of generalizing across different geographic regions and trajectory data mining tasks remains an emergent yet important problem.

In recent years, many efforts have been made to address this problem. These works can be categorized into two groups, \ie trajectory foundation models and large language model (LLM)-based models. Methods in the first group \cite{jiang2023self,yang2023lightpath,xu2025mm} pre-train general trajectory representations on large-scale dataset and employ several learnable task heads for task adaptation. Due to the geographical heterogeneity, the learned representation models are only effective in trained regions and hard to extend to other data-scarce regions. For LLM-based models~\cite{zhou2024plm4traj,yu2024bigcity,zhang2024large,wei2025path}, researchers attempt to utilize the generalization ability of LLMs for trajectory data mining. The LLMs are either treated as well-trained encoders directly feeding raw trajectories to obtain the task results~\cite{zhang2024large} or fine-tuned to enhance traditional trajectory foundation models~\cite{zhou2024plm4traj,yu2024bigcity,wei2025path}. Nevertheless, LLMs struggle with handling numerical data and performing arithmetic operations\cite{li2025stbench}, which limits their performances to encode raw trajectories. The commonly fine-tuned LLM is GPT-2\cite{yu2024bigcity,zhou2024plm4traj}  which has only 1.5B parameters and  exhibits limited generalization capability compared to modern, larger-scale models. Therefore, the generalization problem of trajectory data mining remains unsolved. 

\begin{figure}[t]
	\centering
	\includegraphics[width=0.46\textwidth]{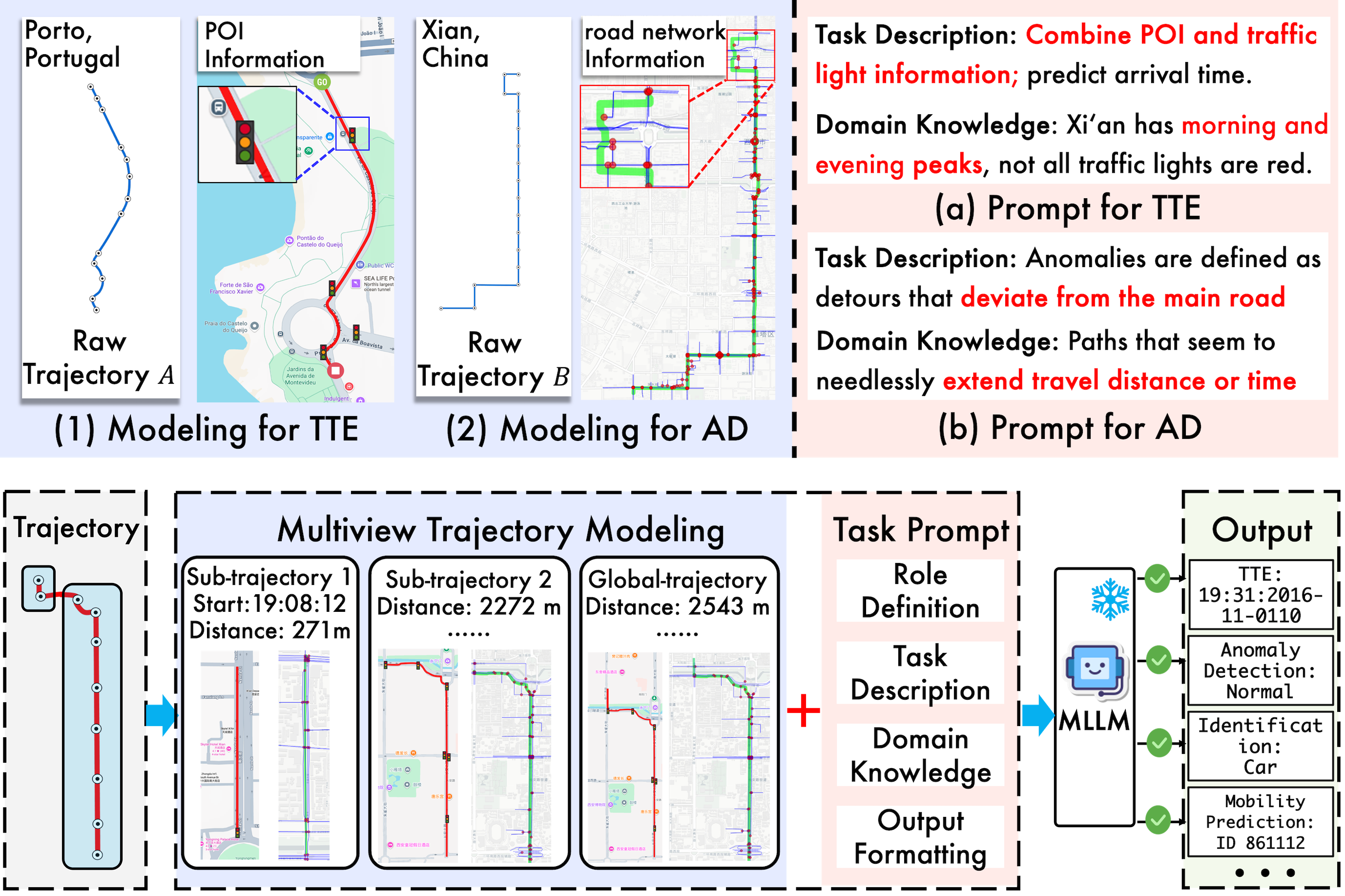}
        \vspace{-0.3cm}
	\caption{The motivation of \our.}
	\label{fig:motivation}
\vspace{-0.65cm}
\end{figure}

Notably, multimodal large language models (MLLMs) have shown remarkable reasoning ability, excelling in integrating multimodal information to enhance the performance of various tasks. 
The recent advances of MLLMs raise the question: can MLLMs reform current trajectory data mining and solve the generalization problem? Raw trajectories, recording as the sequences of GPS coordinates, are the simplified spatial-temporal expression of human movements. By projecting trajectories on the map, visual contextual information, such as POIs, road networks and traffic lights, can be represented in the visual modality. MLLMs' natural language processing (NLP) capability allows us to capture the regional information, human knowledge and task description with text. Moreover, well-trained MLLMs integrate rich common-sense knowledge and have strong generalization ability across different scenarios. It has significant potential to understand trajectories from visual and textual modalities without extra training and extend to analyze trajectories in unseen regions. 

However, due to the modality gap of trajectory, two foundational challenges should be addressed to achieve region-agnostic and task-adaptive trajectory data mining with MLLMs:
\begin{itemize}[leftmargin=4mm]
    \item \textbf{Task-independent Trajectory Modeling.} The spatial-temporal characteristics vary a lot in trajectories and the required contextual information is heterogeneous for different tasks, leading to the challenge for modeling cross-region trajectories in a task-independent way. As shown in Figure~\ref{fig:motivation}, trajectories $A$ and $B$ are in different regions and $A$ is much shorter than $B$. The information required for travel time estimation(TTE) is also different from anomaly detection(AD). Directly compressing all factors in one image would obscure real useful information leading to suboptimal performance. The trajectory modeling should not only capture the heterogeneity of all spatial-temporal and contextual information, but also extract region-agnostic features useful across different geographical regions. 

    \item \textbf{Flexible Task Adaptation.} Instead of using trainable task heads for adaptation~\cite{zhou2024plm4traj}, we argue that textual prompts are more effective to activate the multimodal reasoning ability of MLLMs. The prompts can easily integrate human knowledge and define the reasoning steps for tasks clearly. As illustrated in Figure~\ref{fig:motivation}, the task descriptions, reasoning steps and output formats of TTE and AD are totally different. Nonetheless, these prompts should be easy to reference the information contained in trajectory modeling and data-invariant to meet the requirements of various tasks across different trajectory datasets. Therefore, it is challenging to construct task-specified prompts and achieve flexible task adaptation. 
\end{itemize}

To address these challenges, we propose \our which is the first general framework that employs MLLMs for trajectory data mining. For unified trajectory modeling, we design a map-anchored tokenization mechanism which segments trajectories into sub-trajectories based on semantic completeness on the map. Both the visual and text representations are generated on the granularity of sub-trajectories. Taking these sub-trajectories as inputs, \our constructs different views of spatial scales and contextual factors that contain all required information for different tasks. For example, the trajectory $A$ in Figure~\ref{fig:motivation} is described at global and sub-trajectory scales. The POIs and road networks information are modeled as the images and texts for each scale. Utilizing the multimodal reasoning ability of MLLMs, the temporal constraints are captured with the interleaved image-text sequences of visual and text representations. To achieve flexible task adaptation, we proposed a prompt optimization method that employs few seed trajectories and performs multi-round interactions with MLLMs to finalize the data-invariant task prompts. In usage, an arbitrary trajectory can be transformed into multimodal inputs of MLLMs to generate the analyzing results of various tasks. 

By doing so, \our has three attractive benefits: (1) \textbf{Training-free.} \our relies on the reasoning ability of MLLMs to conduct trajectory data mining and does not need any data for model training or fine-tuning. This allows \our to be suitable for handling trajectories in arbitrary regions. (2) \textbf{Extensible.} Both the trajectory modeling and prompt optimization are general and independent, which makes \our extensible to new tasks by involving other information views and task prompts. (3) \textbf{Interpretable.} As the by-products, \our also outputs the reasoning process, which can be used to interpret the results.

The key contributions of this paper are summarized as follows:
\begin{itemize}[leftmargin=4mm]
    \item We propose a novel framework, namely \our, which employs MLLMs for region-agnostic and task-adaptive trajectory data mining. To the best of our knowledge, this is the first work to employ the reasoning ability of MLLMs to analyze human movement trajectories.

    \item To model trajectories in a unified way, we design a map-anchored mechanism to segment trajectories into sub-trajectories, and represent them in visual and text modalities, considering multiple views. For adapting to different tasks, a prompt optimization method is proposed to finalize data-invariant task prompts.

    \item Extensive experiments on four datasets demonstrate \our achieves significant performance improvements, \ie $48.05\%$, $15.52\%$, $51.52\%$, $1.83\%$ on four representative tasks, compared with state-of-the-art baselines. These results are achieved without any training.

    \item We release a dataset that contains the multimodal responses of four trajectory mining tasks on seven latest MLLM backbones. This dataset can shed light on future trajectory data mining research, such as fine-tuning a new MLLM specially used for trajectory data.

\end{itemize}

\begin{figure*}[t]
\centering
\includegraphics[width=1.0\textwidth]{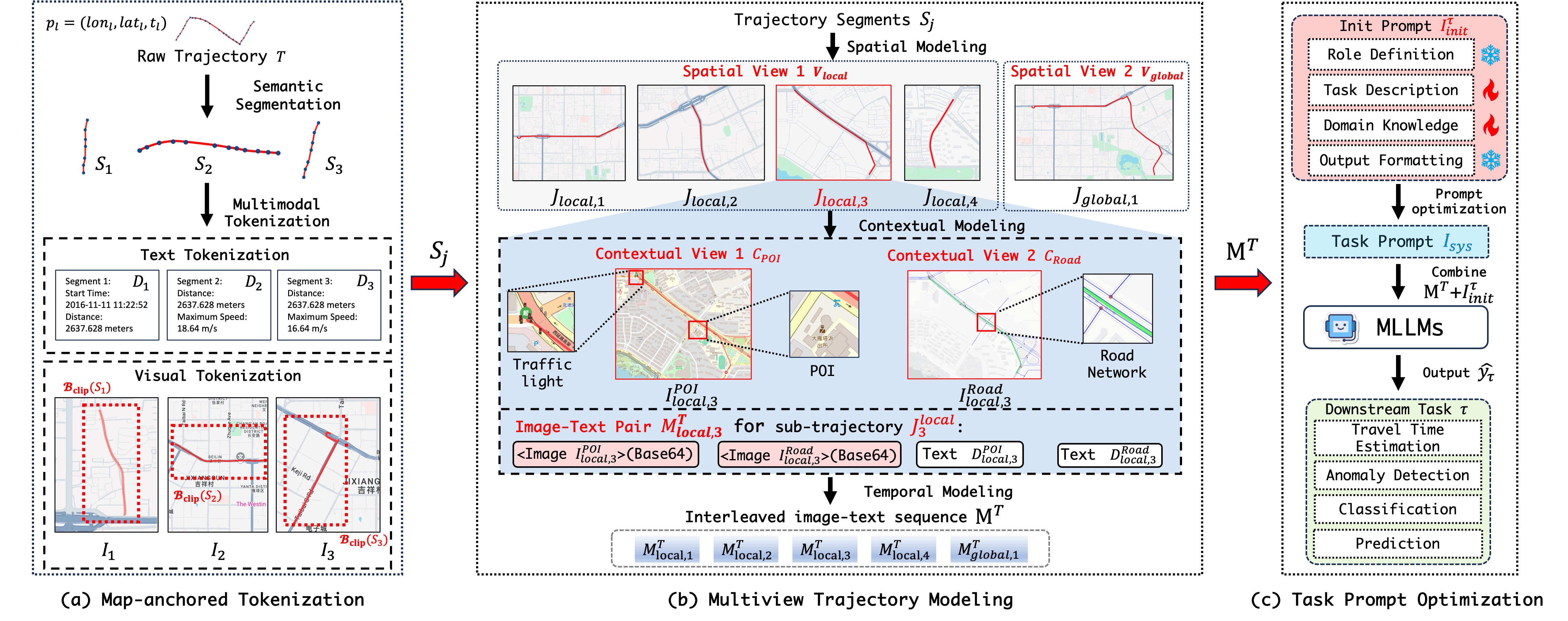}
\vspace{-2ex}
\caption{Overview of \our. Instead of direct model training, \our transforms raw trajectories into interleaved image-text sequences through map-anchored tokenization and multiview trajectory modeling. This enables us to reframe downstream tasks as multimodal reasoning problems, which can be directly solved by MLLMs in a training-free manner.\label{fig:overview}} 
\vspace{-1ex}
\end{figure*}

\section{Related Works}
\label{sec:related}

\noindent\textbf{Trajectory Data Mining.}
Existing research in trajectory data mining can be broadly divided into three categories. Early approaches focus on task-specific models designed for solving one specific task, such as travel time estimation~\cite{jiang2023self,yang2023lightpath,zhou2024plm4traj,xu2025mm}, trajectory prediction~\cite{feng2018deepmove, zhou2024plm4traj}, anomaly detection~\cite{liu2020online, han2022deeptea, wang2024multi}, or classification~\cite{jiang2017trajectorynet,dabiri2019semi,liang2022trajformer}. These works embed assumptions in model design, limiting their generalization ability for other tasks. To address this, trajectory foundation models~\cite{li2018deep,yao2017trajectory,yao2019computing,zhang2020trajectory,yang2021t3s, jiang2023self,yang2023lightpath,zhao2025unitr, zhou2025blue} are proposed to learn a unified model for various tasks. However, large-scale trajectories are required to pretrain these foundation models, leading to suboptimal performance on trajectories of unseen regions. Witnessing the progress of LLMs, several works try to solve the generalization problem with LLMs. These works treat LLMs as encoders~\cite{yu2024bigcity,zhou2024plm4traj,wei2025path}, enhancer~\cite{zhang2024large} and agents~\cite{du2024trajagent} to boost trajectory data mining. Nevertheless, the used LLMs are language models failing to handle image data, which is suitable for modeling the spatial and contextual information of trajectory. These researches inspire us to propose \our which is the first general framework utilizing MLLMs for trajectory data mining.

\noindent\textbf{Multimodal Large Language Models.}
MLLMs have become a hot topic of recent AI research. Numerous works are proposed to either align visual modality to text\cite{radford2021learning,li2022blip,li2023blip} or enhance the multimodal reasoning ability\cite{zheng2023ddcot,kil2024mllm,dong2025insight}. The recent advances of MLLMs raise many geographical applications\cite{vivanco2023geoclip,li2024georeasoner,zhang2024earthgpt,kuckreja2024geochat,xu2024addressclip}. For example, MLLMs are now capable of inferring the geographic location of images, such as those shared on social media or captured from street views, based solely on visual content.\cite{vivanco2023geoclip,li2024georeasoner}
However, most of current works are designed for analyzing static images, \eg remote sensing pictures. Trajectory contains rich spatial-temporal and contextual information. Thus, current works cannot be directly used for trajectory data mining.

\section{Preliminaries}
In this section, we first define the problem of MLLM-based trajectory data mining, followed by a brief overview of \our.

\noindent\textbf{Problem Definition:} 
Assuming that a raw trajectory $T$ consists of a sequence of points, $T = \{p_1, \dots,p_l,\dots, p_L\}$, where each point $p_l$ consists of a GPS coordinate pair and a timestamp, \ie, $(\text{lon}_l, \text{lat}_l, t_l)$. Trajectory data mining tasks are denoted as a set $\mathcal{T}$, where  $\tau \in \mathcal{T}$  is one specific task, such as travel time estimation, anomaly detection, or trajectory prediction. The problem of MLLMs-based trajectory data mining can be formally defined as follows: 

Taking raw trajectory $T$ and a task $\tau$ as inputs, there is a multimodal mapping function $\mathcal{M}$ that first transforms $T$ into visual and text modalities, denoted as $\bm{I}^T$ and $\bm{D}^T$, and combines them to obtain the interleaved image-text sequence $\bm{M}^T$.  There is another prompt construction function $\mathcal{P}$ that generates the task prompt $P^\tau$ for task $\tau$. Note that $\mathcal{M}$ and $\mathcal{P}$ are independent of each other, which makes the problem general for different geographical regions and tasks. The goal of MLLMs-based trajectory data mining is to generate the analysis result $\hat{y}_\tau$ = $\text{MLLM}(\bm{M}^T, P^\tau)$.

\noindent\textbf{Overview of \our:} As illustrated in Figure~\ref{fig:overview}, \our consists of three key modules, \ie, map-anchored tokenization, multiview trajectory modeling, and task prompt optimization. The first two modules are designed to build the multimodal mapping function $\mathcal{M}$, and the third module is proposed for $\mathcal{P}$. To bridge the modality gap of trajectory data, raw trajectory $T$ is segmented into sub-trajectories by map-anchored tokenization which encourages records having complete semantic information in one sub-trajectory. To model all the information required for different tasks, spatial, contextual, and temporal views of these sub-trajectories are constructed in the multiview trajectory modeling module to obtain an interleaved image-text sequence $\bm{M}^T$ of $T$. Finally, the task prompt optimization module is designed to obtain $\mathcal{P}$ and achieve flexible task adaptation. 
\section{Methodology}
In this section, we elaborate the three key modules in \our, \ie, map-anchored tokenization, multiview trajectory modeling and task prompt optimization in Section \ref{sec:tokenization}, \ref{sec:multiview} and  \ref{sec:prompt}, respectively. As the whole method is formulated for one trajectory, we omit the subscript $T$ in the notations of following sections.

\subsection{Map-Anchored Tokenization}\label{sec:tokenization}
Raw trajectory data consists of numerical coordinates and timestamps, lacking clear semantic structure, which makes it difficult to understand by MLLMs directly. Therefore, this module decomposes trajectories into semantic coherent sub-trajectories, and transform them into visual and text tokens.

\subsubsection{Semantic Segmentation}
A trajectory may contain semantically distinct phases, such as high-speed driving on highways, low-speed urban navigation, or stationary periods at intersections. Instead of feeding raw trajectory, sub-trajectories with coherent semantics can better capture fine-grained details and unleash the reasoning power of MLLMs. Therefore, \our conducts semantic segmentation at first and formulate it as an optimization problem.

Given a raw trajectory $T$, semantic segmentation aims to find a globally optimal segmentation $\bm{S} = \{S_1,\dots,S_n,\dots, S_N\}$ by minimizing the sum of segment costs. We first project $T$ on the map and define a cost function $\mathrm{cost}(a, b)$ containing $E$ map-anchored factors to quantify the semantic consistency of sub-trajectory from $p_a$ to $p_b$:
\begin{equation}
    \label{eq:cost}
    \mathrm{cost}(a, b) = \sum_{e=1}^{E} f_e(a,b)
    \end{equation}
In this paper, we utilize three factors, \ie $f_{speed}$, $f_{road}$ and $f_{len}$ that capture motion, routing, and structural properties respectively. 
$f_{speed}(a, b)$ measures speed consistency that encourages sub-trajectory with stable motion, such as cruising at a constant speed or being stationary. $f_{road}(a, b)$ represents the number of road type changes, which prevents a sub-trajectory having different road types,  $f_{len}(a, b)$ is a length regularization term that penalizes breaking the trajectory into too many short sub-trajectories. More details can be found in Appendix~\ref{append:factor}.

To solve the optimization problem, we employ dynamic programming. Let $\mathrm{DP}[h]$ denote the minimum cumulative cost for an optimal segmentation of the prefix trajectory $\{p_1, \dots, p_h\}$. The recurrence relation is defined by:
\begin{equation}
\label{eq:dp}
\mathrm{DP}[h] = \min_{1 \leq d < h} \left( \mathrm{DP}[d] + \mathrm{cost}(d+1, h) \right)
\end{equation}
Here, $h$ denotes the current endpoint of the subtrajectory for which the segmentation cost is being computed, and $d$ indicates the candidate previous segmentation point. By computing $\mathrm{DP}[h]$, we obtain the globally minimal cost for the entire trajectory. Then, by backtracking the DP table, the specific segment boundary $\{S_1, \dots, S_N\}$ that produces the optimal cost can be recovered. This approach effectively balances computational efficiency and segmentation quality, making it suitable for large-scale trajectory data analysis.

\subsubsection{Multimodal Tokenization}\label{sec:multimodal}
After segmenting the trajectory into sub-trajectories, we convert each sub-trajectory $S_n\in \bm{S}$ into multimodal tokens suitable for MLLM inference. Specifically, we generate two complementary representations: structured text tokens ($D_n$) that summarizes meta features and basic statistics; map-projected visual tokens ($I_n$) that combine spatial and contextual information. Here, we only provide a general method to generate these tokens. More details to obtain the interleaved image-text sequence of $T$ are specified in Section~\ref{sec:multiview}. 

\noindent\textbf{Text Tokenization.} 
For each sub-trajectory $S_n$, multiple statistical features are extracted from raw trajectory to generate a structured text description $D_n$. 
\begin{equation}
\label{eq:text_token}
D_n = \text{Concat}\big(\text{Time}(S_n), \text{Dist}(S_n), \text{Speed}(S_n), \dots\big)
\end{equation}
The components here include Time($\cdot$) for the start and end timestamps, Dist($\cdot$) for the total distance traveled, and Speed($\cdot$) for summarizing dynamic metrics such as average speed and maximum speed.
These features are formatted to structured natural language sentences, providing semantic descriptions of $S_n$. More details can be found in Appendix~\ref{append:description}.

\noindent\textbf{Visual Tokenization.} 
To complement the text tokens, we generate visual tokens $I_n$ which is the map-projected image. The image generation process contains three steps:

\noindent(1) \textit{Trajectory Rendering:} Each sub-trajectory $S_n$ is rendered as a visually distinct polyline on a map tile service (e.g., OpenStreetMap). The start and end points of the segment are marked with unique icons to indicate direction.

\noindent(2) \textit{Map Clipping:} To focus the image on the relevant map area while preserving surrounding context, we dynamically calculate a minimal bounding box that encompasses all points within $S_n$. This bounding box is then expanded by a predefined padding factor $\Delta$ to obtain the clipping area. 

\noindent(3) \textit{Image Capturing:} Leveraging a headless browser engine (e.g., Puppeteer), we capture a screenshot of the rendered map within the computed clipping area $\mathcal{B}_{\text{clip}}(S_n)$.

Note that the method above is a general version to clarify the basic idea of multimodal tokenization. Next, we will detail how to integrate spatial, contextual, and temporal information with the tokenization method.
\subsection{Multiview Trajectory Modeling}\label{sec:multiview}
After trajectory tokenization, a core challenge remains: how to uniformly integrate sufficient information for meeting the diverse demands of tasks and model the trajectory in a unified way. 
Given a sequence of sub-trajectories $\bm{S}=\{{S_1,\dots,S_n,\dots,S_N}\}$ of trajectory $T$, multiview trajectory modeling aims to construct a multi-view interleaved image-text sequence $M^T$ which contains different views of spatial scales  $\bm{V}=\{V_1, \dots,V_k,\dots,V_K\}$ and contextual information  $\bm{C}=\{C_1,\dots,C_z,\dots,C_K\}$.


\begin{algorithm}[htbp!]
\caption{Multiview Trajectory Modeling}
\label{alg:multiview_modeling}
\begin{flushleft}
\textbf{Input}: A sequence of sub-trajectories of a trajectory $T$, denoted as $\bm{S}=\{{S_1,\dots,S_n,\dots,S_N}\}$.\\
\textbf{Output}: Multiview interleaved image-text sequence $\bm{M}^T$.
\end{flushleft}

\begin{algorithmic}[1]
\STATE $\bm{M}^T \leftarrow []$
\STATE $\bm{V}=\{V_1,\dots,V_K\} \leftarrow \text{GenerateSpatialViews}(\bm{S},\mathcal{G})$
\STATE $\bm{C}=\{C_1,\dots,C_K\} \leftarrow \text{GenerateContextualViews()}$ 
\FOR{each spatial view $V_k \in \bm{V}$}
    \FOR{each sub-trajectory $J_{k,i} \in J_k$}
        \FOR{each contextual view $C_z \in \bm{C}$}
            \STATE $I_{k,i}^z \leftarrow \text{GenerateVisual}(J_{k,i}, C_z)$ \COMMENT{Using the context information provided by $C_j$, generate image based on view $J_{k,i}$}
            \STATE $D_{k,i}^z \leftarrow \text{GenerateText}(J_{k,i}, C_z)$ \COMMENT{Generate statistical text description for $J_{k,i}$}
            \STATE $\bm{M}^T.\text{append}((I_{k,i}^z, D_{k,i}^z))$
        \ENDFOR
    \ENDFOR
    
\ENDFOR
\RETURN $\bm{M}^T$
\end{algorithmic}
\end{algorithm}
\vspace{-3ex}
\subsubsection{Spatial Information Modeling}
To model coarse-to-fine spatial information of $T$, a set of $K$ spatial views, $\bm{V}=\{V_1, \dots,V_k,\dots,V_K\}$ is constructed to describe the trajectory at different scales. Each spatial view $V_k$ represents a scale of trajectory $T$ and can be obtained by combining sub-trajectories with a specific coarse-grained strategy $g_k$. For trajectory $T$, we construct a set of coarse-grained strategies $\mathcal{G} = \{{g_1, \dots, g_k, \dots, g_K}\}$ which convert multimodal representations for different spatial scales.

Formally, by applying $g_k$ on the sub-trajectories $\bm{S}$, we can obtain a partition of $\bm{S}$, denote as $J_k = \{J_{k,1}, \dots, J_{k,i}, \dots, J_{k,I}\}$, where $J_{k,i}$ is a combination of successive sub-trajectories, \eg, $J_{k,1} = \{S_1, S_2, \dots \}$ and $J_{k,i} = \{S_{n-1}, S_n, \dots\}$. As illustrated in Figure~\ref{fig:overview}(b), we construct two coarse-grained strategies, \ie, $g_{global}$ and $g_{local}$. $g_{global}$ takes $T$ as a whole to generate the global view $V_{global}$, and $g_{local}$ treats each sub-trajectory independently to generate the local view $V_{local}$. Notice that $T$ contains four sub-trajectories, the spatial information modeling of $T$ can be formalized as follows: 
 $V_{local} = \{J_{{local},1},J_{{local},2},J_{{local},3}, J_{{local},4}\}$, and $V_{global} = J_{{global},1}$.

To obtain multimodal representation, each $J_{k,i}$ can also be viewed as a sub-trajectory of $T$. We can get the structured text tokens $D_{k,i}$ and visual tokens $I_{k,i}$ for each sub-trajectory $J_{k,i}$ by applying the \textbf{Map-Anchored Tokenization} module described in Section~\ref{sec:tokenization}. The combination of multimodal representations of these spatial views captures the coarse-to-fine spatial semantics of the trajectory.






\subsubsection{Contextual Information Modeling}
\our enriches the visual tokens of spatial view partitions to integrate various contextual information. We introduce a set of contextual views $\bm{C} = \{C_1, \dots, C_z, \dots, C_Z\}$, each representing a different type of contextual information relevant to the trajectory, such as POIs, road networks, and traffic lights. 
Given an arbitrary $J_{k,i}\in J_k$, we extend the visual tokenization method described in Section~\ref{sec:multimodal} and generate image $\{I_{k,i}^z\}_{z\in 1 ... Z}$ for each context view $C_z \in \bm{C}$. 

As illustrated in Figure~\ref{fig:overview}(b), we enrich the spatial views by POIs and road networks, and obtain $I_{k,i}^{POI}$ and $I_{k,i}^{road}$. We employ two key strategies to ensure the resulting images are informative yet clear.

\noindent\textbf{Layer Decomposition.} Directly overlaying all available contextual data on a single image would obscure the useful information in visual clutter. To address this, we decompose contextual information into distinct visual map layers. Each map layer can be treated as a contextual view of $T$. Thus, the map layer decomposition produces multiple contextual views for a single spatial $J_{k,i}$. As shown, $I_{k,i}^{POI}$ and $I_{k,i}^{road}$ are generated for $T$.

\noindent\textbf{Context Filtering.} However, due to the large amount of contextual data available in a region, it is necessary to filter the most relevant information for the trajectory. Specifically, we filter the context elements rendered in each view based on their relevance to the sub-trajectory $J_{k,i}$. Let $\mathcal{E}_{k,i}^{z}$ denote the full set of context elements associated with view type $C_z$. We define a general scoring function $\text{dis}(e, J_{k,i})$ that measures how closely a context element $e \in \mathcal{E}_{k,i}^{z}$ relates to the sub-trajectory represented in $J_{k,i}$. We retain only those elements below a threshold $\theta$ which represents the relevance of the trajectory:
\begin{equation}
\label{eq:relevance}
\mathcal{E}^{z*}_{k,i} = \{e \in \mathcal{E}_{k,i}^{z} \mid \text{dis}(e, J_{k,i}) \le \theta\}
\end{equation}
In practice, we realize the relevance function primarily based on spatial proximity. For example, in a POI-centric view, the element set $\mathcal{E}_{k,i}^{\text{POI}}$ consists of all POIs in the region. The filtered subset is defined as:
\begin{equation}
\label{eq:poi_filter}
\mathcal{E}^{\text{POI}*}_{k,i} = \{e \in \mathcal{E}^{\text{POI}}_{k,i} \mid \text{dis}^{POI}(e, J_{k,i}) \le \theta_{\text{POI}}\},
\end{equation}
where $\text{dis}^{POI}(e, J_{k,i})$ is the shortest Euclidean distance from the POI $e$ to $J_{k,i}$ and the $\theta_{\text{POI}}$ is the distance threshold.

Furthermore, to visually highlight the trajectories, we present them clearly enough to distinguish trajectories from the contextual data. As shown in Figure~\ref{fig:overview}, we use thick red lines in the POI-centric view or thick green lines in the road network view. These vision and filtering strategies can effectively guide MLLMs to understand critical contextual cues.

\subsubsection{Temporal Information Modeling}
Temporal order is a fundamental property of trajectories. Instead of using a trainable temporal encoder, \our exploits the sequential modeling capability of MLLMs to capture the temporal information. By arranging the multimodal tokens of all generated spatial and contextual views, we can obtain an interleaved image-text sequence for $T$. 

Formally, given a set of multimodal tuples $\{(I_{k,i}^z, D_{k,i}^z)\}$ where $k \in (1, K)$, $i \in [1,I]$ and $z \in [1,Z]$, we assemble them into an ordered interleaved image-text sequence $\bm{M}^T = [M^T_1, \dots,M^T_k,\dots, M^T_K]$ to capture temporal dependencies. Each $M^T_k$ corresponds to a spatial view $V_k$. Each $M^T_k$ is further composed of $I$ spatial sub-views, denoted as $M^T_k = [M^T_{k,1}, \dots,M^T_{k,i},\dots, M^T_{k,I}]$. Each sub-view $M^T_{k,i}$ consists of $Z$ multimodal pairs derived from different contextual views $\bm{C} = \{C_1, \dots,C_z,\dots, C_Z\}$. The overall process is summarized in Algorithm~\ref{alg:multiview_modeling}.


\begin{equation}
\label{eq:subview}
M^T_{k,i} = \big[(I^{1}_{k,i}, D^{1}_{k,i}), \dots, (I^{z}_{k,i}, D^{z}_{k,i})\big],
\end{equation}
where $(I^{z}_{k,i}, D^{z}_{k,i})$ denotes the image-text pair in contextual view $C_z$.

As shown in Figure~\ref{fig:overview}(b), the sequence $M^T_{\text{local},3}$ refers to the third sub-trajectory $J_{local,3}$ of the local spatial view $V_{local}$. It contains multimodal pairs from two contextual views: the POI-centric view $C_{\text{POI}}$ and the road network view $C_{\text{road}}$, denoted as $(I^{\text{POI}}_{\text{local},3}, D^{\text{POI}}_{\text{local},3})$ and $(I^{\text{road}}_{\text{local},3}, D^{\text{road}}_{\text{local},3})$, respectively.

The images are encoded as base64 to fit the inputs of MLLM prompt. The corresponding text description both provides a semantic interpretation of the image and serves as a text anchor (e.g., "POI image of segment 1: <image>"). This structured and sequential presentation enables MLLMs to naturally interpret trajectories as temporally coherent narratives, effectively capturing time-dependent relationships without additional training.


\subsection{Task Prompt Optimization}\label{sec:prompt}
To achieve flexible task adaptation, we propose to model the task-specified information in the task prompt instead of training task heads. However, both the reasoning steps and used information of different tasks vary significantly. It is challenging to construct a data-invariant task prompt that can be easily understood by MLLMs and suitable for different trajectories. This module aims to address this challenge with a few labeled seed trajectories (e.g., less than 10) and conducts multi-rounds interaction with MLLMs to optimize the task prompts. Next, we first introduce the construction of task prompt template, and then describe the optimization process.

\begin{figure}[t]
\centering
\includegraphics[width=0.4\textwidth]{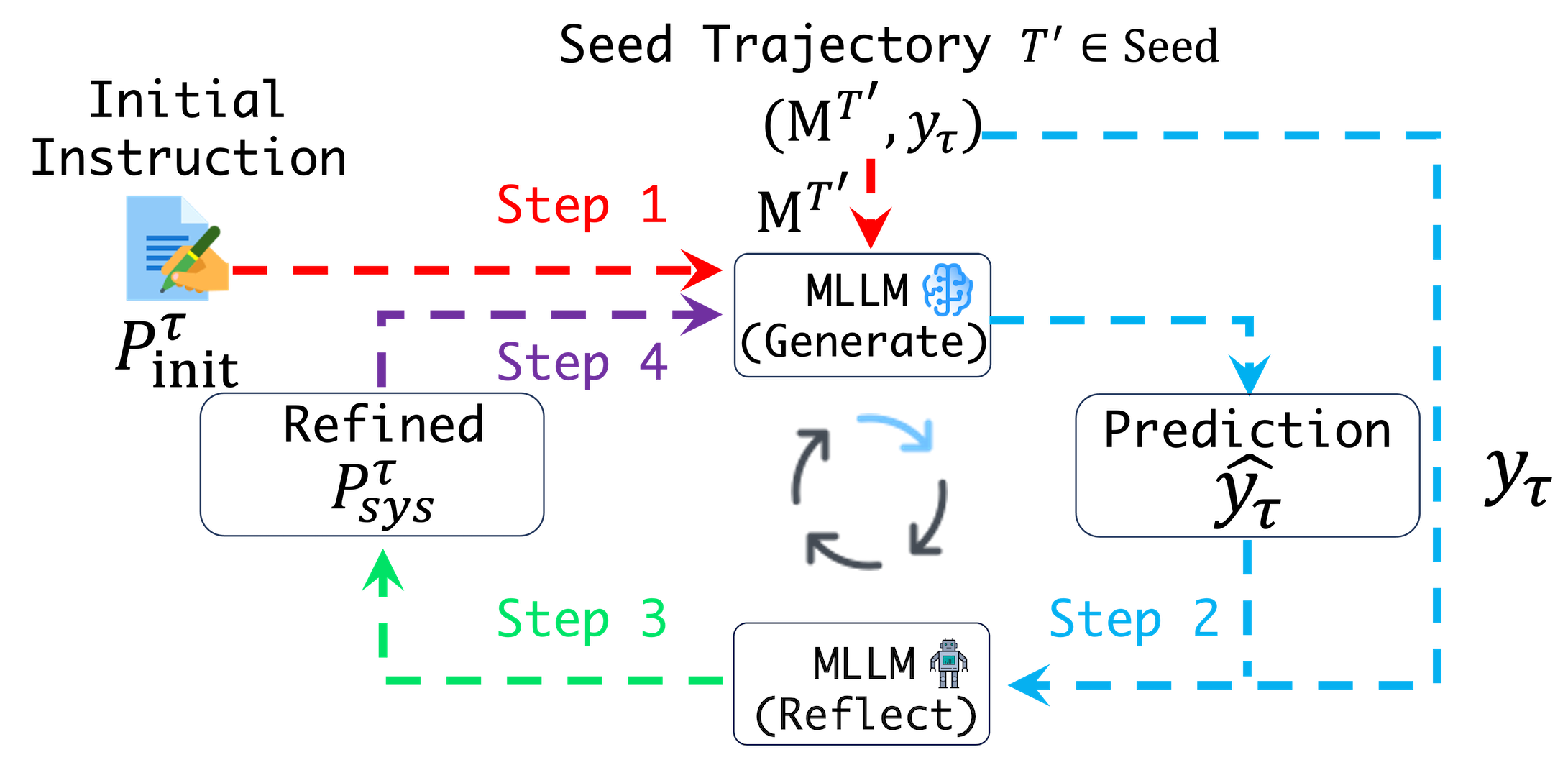}

\caption{Task Prompt Optimization.\label{fig:taskprompt}} 
\vspace{-3ex}
\end{figure}

\subsubsection{Task Prompt Template}
For each trajectory mining task $\tau$, we generate the task prompt with a general prompt template. The task prompt $P_{\text{sys}}$ consists of four components:
\begin{equation}
\label{eq:sys_prompt}
P_{\text{sys}} = (P_{\text{role}}, P_{\text{task}}, P_{\text{know}}, P_{\text{format}})
\end{equation}
Here, the Role Definition ($P_{\text{role}}$) and Output Format ($P_{\text{format}}$) are fixed for each task. The optimization focuses on refining the Task Description ($P_{\text{task}}$) and Domain Knowledge ($P_{\text{know}}$). 

\subsubsection{Prompt Optimization}
In this module, \our takes an initial prompt $P_{\text{init}}^{\tau}$ predefined by human experience and a few seed trajectories as inputs. We aim to optimize $P_{\text{init}}^{\tau}$ into a more effective prompt $P_{\text{sys}}^{\tau}$ through an automated, MLLM refinement loop. As shown in Figure \ref{fig:taskprompt}, the optimization proceeds as follows: 
(1) We first select a set of seed trajectories $\bm{Seed}$ and generate their corresponding interleaved image-text sequence set $\mathbb{M} = \{\bm{M}^{T'} \mid T' \in \ \mathcal{S}\}$;
(2) For each $T' \in \bm{Seed}$, we feed its interleaved image-text sequence $\bm{M}^{T'}$ and the initial system prompt $P_{\text{init}}^{\tau}$ into the MLLM to obtain the output $y$;
(3) We construct a feedback loop, \ie, MLLM reflects on its output $y$ in light of the true label $\hat{y}$, generating suggestions to improve the prompt. Based on the feedback from disagreement samples, the MLLM generates a refined prompt $P_{\text{sys}}^{\tau}$, which is subsequently used to re-evaluate other seed trajectories;
(4) Repeat (2)-(3) several times, allowing MLLM to self-improve the prompt based on feedback. The process concludes when the refined prompt $P_{\text{sys}}^{\tau}$ demonstrates satisfactory performance on $\bm{Seed}$.

Notice that once $P_{\text{sys}}^{\tau}$ is finalized, it can be directly reused across different regions by simply updating some location-specific parts in $P_{\text{know}}$, such as the population and city scale.

\textbf{Model Inference.} Once $P_{\text{sys}}^{\tau}$ is finalized, inferring any new trajectory $T$ is straightforward. The trajectory is first transformed into an interleaved image-text sequence $\bm{M}^{T}$ (see Sections~\ref{sec:tokenization} and~\ref{sec:multiview}). This sequence, together with the optimized prompt $P_{\text{sys}}^{\tau}$, is used as input to the MLLM, which generates a textual response. The output is then parsed according to $P_{\text{format}}$ to extract structured results.

\begin{table*}[htbp]
\centering
\small
\renewcommand{\arraystretch}{0.88}
{
\setlength{\tabcolsep}{3pt} 
\caption{Performance on Travel Time Estimation and Mobility Prediction.}
\vspace{-2ex}
\label{tab:perf_merged_final}
\begin{tabular}{|l|ccc|cc|ccc|cc|ccc|cc|}
\hline
\multirow{3}{*}{\textbf{Methods}} & \multicolumn{5}{c|}{\textbf{Xian}} & \multicolumn{5}{c|}{\textbf{Porto}} & \multicolumn{5}{c|}{\textbf{Chengdu}} \\ \cline{2-16}
& \multicolumn{3}{c|}{Travel Time Est.} & \multicolumn{2}{c|}{Mobility Pred.} & \multicolumn{3}{c|}{Travel Time Est.} & \multicolumn{2}{c|}{Mobility Pred.} & \multicolumn{3}{c|}{Travel Time Est.} & \multicolumn{2}{c|}{Mobility Pred.} \\ \cline{2-16}
& MAE & RMSE & MAPE & ACC@1 & ACC@5 & MAE & RMSE & MAPE & ACC@1 & ACC@5 & MAE & RMSE & MAPE & ACC@1 & ACC@5 \\ \hline

Traj2Vec\cite{yao2017trajectory} & 230.98 & 382.31 & 28.18 & 42.31 & 72.99 & 210.12 & 362.76 & 26.61 & 43.77 & 73.98 & 238.92 & 390.11 & 29.11 & 42.12 & 71.88 \\
T2Vec\cite{li2018deep} & 228.84 & 377.63 & 27.28 & 43.21 & 74.01 & 208.22 & 355.47 & 26.23 & 44.22 & 75.27 & 234.52 & 378.24 & 28.87 & 44.01 & 73.23 \\
Trember\cite{fu2020trembr} & 223.73 & 377.23 & 25.82 & 49.72 & 79.17 & 204.49 & 347.66 & 24.56 & 49.57 & 79.82 & 231.55 & 377.15 & 25.92 & 49.72 & 79.54 \\
CTLE\cite{lin2021pre} & 232.29 & 377.34 & 26.82 & 50.83 & 80.14 & 205.53 & 355.82 & 26.25 & 50.22 & 80.65 & 228.52 & 388.31 & 25.88 & 50.23 & 81.02 \\
Toast\cite{chen2021robust} & 226.65 & 378.92 & 26.18 & 50.57 & 80.07 & 204.24 & 353.43 & 26.11 & 50.09 & 80.21 & 228.12 & 386.78 & 25.81 & 49.97 & 80.22 \\
TrajCL\cite{chang2023contrastive} & 218.12 & 365.12 & 25.44 & 49.81 & 79.92 & 198.23 & 340.44 & 24.11 & 49.96 & 79.81 & 222.13 & 368.33 & 25.65 & 49.82 & 79.87 \\
START\cite{jiang2023self} & 203.11 & 339.90 & 24.01 & 49.71 & 80.29 & 195.49 & 354.22 & 23.84 & 51.98 & 81.37 & 196.44 & 314.14 & 24.12 & 53.22 & 86.47 \\
MMTEC\cite{lin2023pre} & 201.23 & 332.18 & 23.91 & 52.25 & 81.16 & 194.21 & 352.34 & 23.67 & 51.99 & 81.13 & 195.32 & 318.98 & 24.02 & 53.12 & \underline{86.52} \\
UniTR\cite{zhao2025unitr} & 204.93 & 345.33 & 25.14 & 51.67 & 79.39 & 199.41 & 347.59 & 23.82 & 50.11 & 80.14 & 208.44 & 339.23 & 24.82 & 52.87 & 84.10 \\
JGRM\cite{ma2024more} & 202.19 & 332.88 & 23.98 & 52.28 & 81.02 & 194.43 & 352.45 & 23.78 & 52.01 & 81.29 & 195.34 & 314.62 & 24.01 & 53.11 & 86.42 \\
BLUE\cite{zhou2025blue} & \underline{186.34} & \underline{298.13} & \underline{19.35} & 49.45 & 80.44 & \underline{171.22} & \underline{298.11} & \underline{19.32} & 48.43 & \underline{81.72} & \underline{169.22} & \underline{264.93} & \underline{20.11} & 57.90 & 83.24 \\
MM-Path\cite{xu2025mm} & 197.41 & 317.85 & 22.85 & 48.76 & 81.67 & 181.16 & 320.66 & 22.12 & 48.16 & 80.92 & 187.03 & 279.40 & 22.76 & 58.88 & 84.17 \\
BigCity\cite{yu2024bigcity} & 198.78 & 325.98 & 23.69 & 50.12 & 80.33 & 186.91 & 329.39 & 22.72 & 51.44 & 81.01 & 188.42 & 301.11 & 23.18 & 55.23 & 81.76 \\
PLM4Traj\cite{zhou2024plm4traj} & 196.72 & 334.42 & 23.61 & \underline{52.76} & \underline{82.15} & 189.99 & 344.64 & 22.89 & \underline{52.33} & 81.44 & 189.09 & 298.24 & 23.21 & \underline{59.50} & 86.27 \\
Path-LLM\cite{wei2025path} & 192.76 & 313.88 & 23.54 & 50.85 & 80.73 & 184.42 & 323.83 & 22.67 & 50.15 & 80.87 & 184.67 & 294.55 & 22.97 & 56.44 & 82.45 \\
\hline
\textbf{\our} & \textbf{131.02} & \textbf{154.86} & \textbf{15.36} & \textbf{60.72} & \textbf{83.25} & \textbf{144.63} & \textbf{173.54} & \textbf{15.14} & \textbf{60.45} & \textbf{82.33} & \textbf{133.43} & \textbf{154.67} & \textbf{16.34} & \textbf{60.12} & \textbf{87.22} \\ \hline
\textit{Improvement} & \textit{29.69\%} & \textit{48.05\%} & \textit{20.62\%} & \textit{15.09\%} & \textit{1.34\%} & \textit{15.53\%} & \textit{41.79\%} & \textit{21.64\%} & \textit{15.52\%} & \textit{0.75\%} & \textit{21.15\%} & \textit{41.62\%} & \textit{18.75\%} & \textit{1.04\%} & \textit{0.81\%} \\ \hline
\end{tabular}
}
\end{table*}

\begin{table*}[htbp]
\centering
\caption{Performance on Anomaly Detection (PR-AUC score).}
\vspace{-2ex}
\label{tab:anomaly_detection}
{
\small
\renewcommand{\arraystretch}{0.88}
\setlength{\tabcolsep}{3.6pt} 
\begin{tabular}{|l|ccc|cc|ccc|cc|ccc|cc|}
\hline
\multirow{3}{*}{\textbf{Method}} & \multicolumn{5}{c|}{\textbf{Xian}} & \multicolumn{5}{c|}{\textbf{Porto}} & \multicolumn{5}{c|}{\textbf{Chengdu}} \\ \cline{2-16}
& \multicolumn{3}{c|}{Detour} & \multicolumn{2}{c|}{Switch} & \multicolumn{3}{c|}{Detour} & \multicolumn{2}{c|}{Switch} & \multicolumn{3}{c|}{Detour} & \multicolumn{2}{c|}{Switch} \\ \cline{2-16}
& Low & Medium & High & $\mu=0.3$ & $\mu=0.5$ & Low & Medium & High & $\mu=0.3$ & $\mu=0.5$ & Low & Medium & High & $\mu=0.3$ & $\mu=0.5$ \\ \hline
iBAT\cite{zhang2011ibat} & 0.439 & 0.487 & 0.694 & 0.199 & 0.262 & 0.273 & 0.313 & 0.501 & 0.201 & 0.241 & 0.450 & 0.497 & 0.691 & 0.201 & 0.254 \\
GM-VSAE\cite{liu2020online} & 0.854 & 0.921 & 0.944 & 0.521 & 0.612 & 0.374 & 0.722 & 0.910 & 0.871 & 0.893 & 0.858 & 0.919 & 0.947 & 0.547 & 0.609 \\
ATROM\cite{gao2023open} & 0.816 & 0.897 & 0.901 & 0.532 & 0.601 & 0.386 & 0.709 & 0.882 & 0.845 & 0.881 & 0.841 & 0.872 & 0.921 & 0.535 & 0.611 \\
DeepTEA\cite{han2022deeptea} & 0.871 & 0.932 & 0.940 & 0.576 & 0.654 & 0.408 & 0.736 & 0.921 & 0.861 & 0.872 & 0.875 & 0.931 & 0.935 & 0.576 & 0.655 \\
MST-OATD\cite{wang2024multi} & 0.902 & 0.951 & 0.971 & 0.598 & 0.691 & 0.739 & 0.892 & 0.970 & 0.899 & 0.912 & 0.907 & 0.949 & 0.969 & 0.604 & \underline{0.677} \\
FOTraj\cite{shao2025fo} & \underline{0.911} & \underline{0.954} & \underline{0.979} & \underline{0.625} & \underline{0.733} & \underline{0.741} & \underline{0.901} & \underline{0.974} & \underline{0.927} & \underline{0.946} & \underline{0.913} & \underline{0.954} & \underline{0.971} & \underline{0.649} & 0.656 \\ \hline
\textbf{\our} & \textbf{0.967} & \textbf{0.979} & \textbf{0.981} & \textbf{0.947} & \textbf{0.956} & \textbf{0.968} & \textbf{0.980} & \textbf{0.981} & \textbf{0.971} & \textbf{0.981} & \textbf{0.967} & \textbf{0.976} & \textbf{0.980} & \textbf{0.966} & \textbf{0.983} \\ \hline
\textit{Improvement} & \textit{6.15\%} & \textit{2.62\%} & \textit{0.20\%} & \textit{51.52\%} & \textit{30.42\%} & \textit{30.63\%} & \textit{8.77\%} & \textit{0.72\%} & \textit{4.75\%} & \textit{3.70\%} & \textit{5.91\%} & \textit{2.31\%} & \textit{0.93\%} & \textit{48.84\%} & \textit{45.19\%} \\ \hline
\end{tabular}
\vspace{-1ex}
}
\end{table*}

\begin{table}[htbp]
\centering
\caption{Performance on Transportation Mode Identification.}
\vspace{-2ex}
\label{tab:traffic_mode_classification_fullwidth}
{
\small
\renewcommand{\arraystretch}{0.88}
\setlength{\tabcolsep}{10.9pt}
\begin{tabular}{|l|ccc|}
\hline
\multirow{2}{*}{\textbf{Method}} & \multicolumn{3}{c|}{\textbf{Geolife}} \\ \cline{2-4}
& Accuracy & Macro-F1 & Weighted-F1 \\ \hline
SECA\cite{dabiri2019semi} & 78.91 & 78.78 & 79.02 \\
TrajODE\cite{liang2021modeling} & 85.48 & 84.98 & 85.17 \\
TrajFormer\cite{liang2022trajformer} & 85.56 & 85.12 & 85.71 \\
RED\cite{zhou2024red} & \underline{85.89} & \underline{85.23} & \underline{86.11} \\ \hline
\textbf{\our} & \textbf{87.46} & \textbf{86.71} & \textbf{87.45} \\ \hline
\textit{Improvement} & \textit{1.83\%} & \textit{1.74\%} & \textit{1.56\%} \\ \hline
\end{tabular}
}
\vspace{-4ex}
\end{table}

\section{Experiment}\label{sec:experiment}
We conduct experiments to answer the following questions:\\
\noindent\textbf{RQ1:} How does \our perform on different tasks?\\
\noindent\textbf{RQ2:} How does \our perform on different MLLM backbones?\\
\noindent\textbf{RQ3:} What is the efficiency of \our?\\
\noindent\textbf{RQ4:} What are the influences of the proposed components?\\
\noindent\textbf{RQ5:} How is the reasoning ability of \our?\\
\noindent\textbf{RQ6:} What are the influences of hyper-parameter $\theta$?


\subsection{Experimental Settings}
In this section, we briefly introduce the datasets, tasks, evaluation metrics, and baselines. All the datasets and codes of \our are released \footnote{https://anonymous.4open.science/r/Traj-MLLM/}. More detailed settings can be found in Appendix~\ref{append:settings}.

\noindent\textbf{Datasets.}
We conduct extensive experiments on four public trajectory datasets, Xian, Chengdu\footnote{https://outreach.didichuxing.com/}, Porto\footnote{https://www.kaggle.com/c/pkdd-15-predict-taxi-service-trajectory} and Geolife\cite{zheng2011geolife}. The map layers, road networks and POIs are obtained by OpenStreetMap\footnote{https://www.openstreetmap.org/}. Due to budget constraints, we select over $80,000$ trajectories and perform four tasks to obtain about $320,000$ MLLMs responses. 

\noindent\textbf{Tasks and Evaluation Metrics.}
We choose four tasks that are widely studied in trajectory data mining, \ie, travel time estimation (TTE)~\cite{xu2025mm,zhou2025blue}, anomaly detection (AD)~\cite{wang2024multi,shao2025fo}, mobility prediction (MP)~\cite{zhou2024plm4traj} and transportation mode identification (TMI)~\cite{zhou2024red}, to evaluate the effectiveness of \our. We follow existing works\cite{xu2025mm,wang2024multi,zhou2024red} to choose the same metrics. For TTE, we report mean absolute error (MAE), root mean squared error (RMSE), and mean absolute percentage error (MAPE). For AD, we evaluate the performance with precision-recall
area under the curve (PR-AUC). For MP, we choose Top-1 Accuracy (ACC@1) and Top-5 Accuracy (ACC@5). For TMI, we report Accuracy, Macro-F1, and Weighted-F1. More details can be found in Appendix~\ref{append:task}.

\noindent\textbf{Baselines.} For each task, we choose three kinds of baselines, \ie, task-specified models, trajectory foundation models and LLM-based models. The baselines for TTE and MP include Traj2Vec~\cite{yao2017trajectory}, T2Vec~\cite{li2018deep}, Trember~\cite{fu2020trembr}, CTLE~\cite{lin2021pre}, Toast~\cite{chen2021robust}, TrajCL~\cite{chang2023contrastive}, MMTEC~\cite{lin2023pre}, JGRM~\cite{ma2024more}, START~\cite{jiang2023self},  UniTR\cite{zhao2025unitr}, BigCity\cite{yu2024bigcity} and BLUE\cite{zhou2025blue}, PLM4Traj~\cite{zhou2024plm4traj} and Path-LLM\cite{wei2025path} and MM-Path~\cite{xu2025mm}. 
For AD, we compare \our against iBAT~\cite{zhang2011ibat}, GM-VSAE~\cite{liu2020online}, ATROM~\cite{gao2023open}, DeepTEA~\cite{han2022deeptea}, MST-OATD~\cite{wang2024multi}, and FOTraj~\cite{shao2025fo}. 
For TMI, \our is compared with SECA~\cite{dabiri2019semi}, TrajODE~\cite{liang2021modeling}, TrajFormer~\cite{liang2022trajformer}, and RED~\cite{zhou2024red}.

\vspace{-3ex}
\subsection{Effectiveness of \our (RQ1)}
To answer \textbf{RQ1}, we compare \our against multiple baselines across four tasks on four datasets. The results show that \our consistently achieves the best performance on all tasks and regions. Notice that \our achieves these results without any training or fine-tuning. 

For TTE, we report the results in Table~\ref{tab:perf_merged_final} and derive three observations. (1) \our outperforms all baselines across all three datasets. For instance, compared to the strongest multimodal baseline MM-Path, \our reduces the MAE by over $29\%$ on Xian, $15\%$ on Porto, and $21\%$ on Chengdu. This consistent improvement demonstrates the effectiveness of \our. (2) Although MM-Path considers the image modality, it is also inferior to \our because multiview trajectory modeling provides richer contextual information to inspire the reasoning ability of MLLMs. (3) \our achieves over $34.9\%$ performance improvements on MAPE compared to the LLM-based methods, \ie PLM4Traj and Path-LLM. This result demonstrates that \our performs better than fine-tuning a small LLM, \eg, GPT-2.

The results of MP are summarized in Table~\ref{tab:perf_merged_final}. MP requires capturing fine-grained spatial-temporal information of the trajectory. Similar to TTE, we observe over $15\%$ improvements of ACC@1 on Xian and Porto compared to the strongest baseline PLM4Traj, demonstrating that \our can guide MLLMs to understand complex and fine-grained mobility patterns.

The results of anomaly detection are presented in Table \ref{tab:anomaly_detection}. From these results, we observe: (1)  \our achieves the best performance among all baselines, consistently outperforming on all datasets, anomaly types, and perturbation levels. (2) \our is robust on both subtle detour and complex switch anomalies, maintaining stable performance under varying perturbation magnitudes. For instance, the PR-AUC score increases from 0.741 to 0.968 for low–magnitude detour anomalies. This improvement can be attributed to the \textit{multiview trajectory modeling}, which provides multi-scale spatial information that encourages the model to capture diverse anomalous patterns. (3) \our achieves high performance without requiring any model training or explicit anomaly examples, enabling to generalize \our to unseen regions.

Moreover, we present the transportation mode identification results in Table~\ref{tab:traffic_mode_classification_fullwidth}. As illustrated,  \our improves accuracy by $1.83\%$ over RED\cite{zhou2024red}. By feeding multimodal information to the MLLM, \our is able to combine visual cues from the map (e.g., the presence of railway tracks) and statistical information from the text prompts (e.g., velocity and acceleration) to infer the transportation mode. In summary, across the four tasks, \our can leverage the MLLM\'s capacity for both fine-grained spatial analysis (as in TTE and MP) and high-level semantic reasoning (as in AD and TMI). according to these results, we argue that \our has strong generalization ability to achieve region-agnostic and task-adaptive trajectory data mining without the requirement of any data.

\subsection{Performance of different MLLMs (RQ2)}
\begin{figure}[t]
	\centering
	\includegraphics[width=0.47\textwidth]{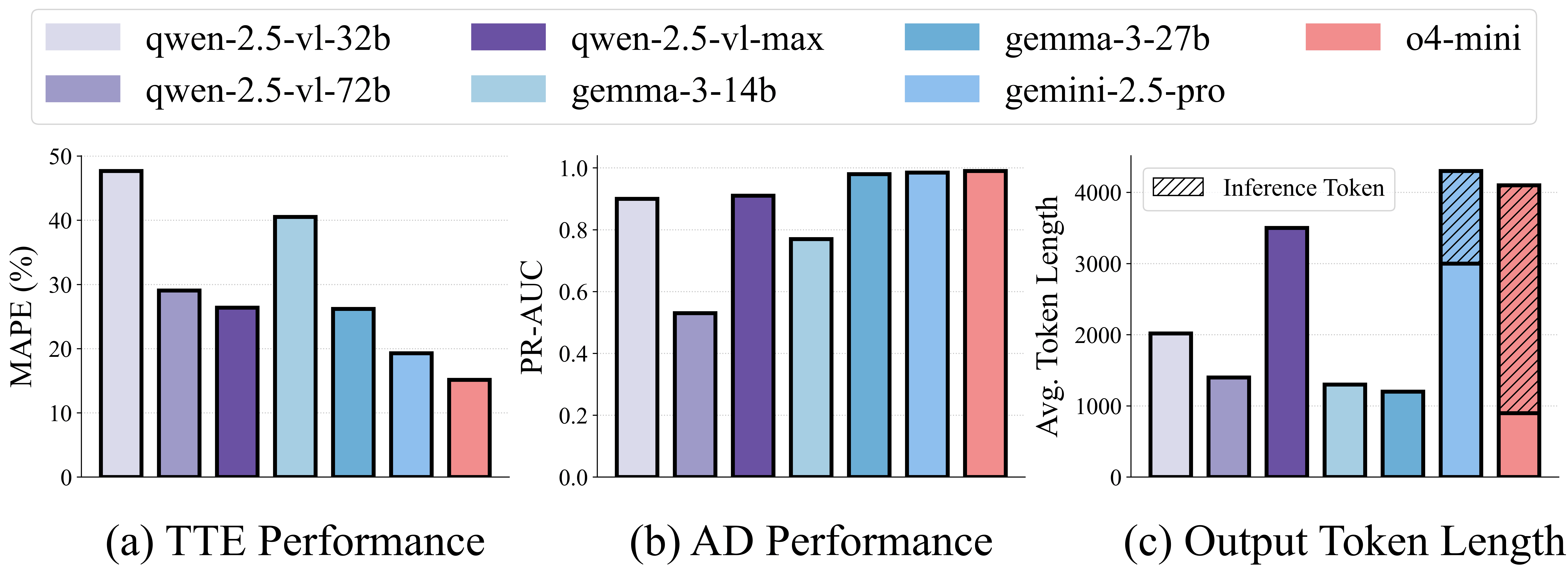}
    \vspace{-1ex}
\caption{Performance with different MLLMs.}	
\vspace{-2ex}
\label{fig:llm_comparison_porto}
\end{figure}

To answer \textbf{RQ2}, we explore the performance of \our under seven MLLM backbones on the Porto dataset across two tasks, \ie TTE and AD (detour anomalies). The results are shown in Figure~\ref{fig:llm_comparison_porto}. According to the results, we have three observations: (1) Within the same model family, closed-source models tend to outperform open-source models. For example, the closed-source qwen-vl-max achieves better results on both tasks than the open-source model qwen-2.5-vl-72b, which may be due to its architecture or proprietary optimizations during training. (2) Model scale is not the sole determinant of performance. Notably, the open-source gemma-3-27b-it performs better than the larger qwen-2.5-vl-72b and achieves a PR-AUC score nearly on par with the powerful closed-source gemini-2.5-pro. This highlights the significant potential and bright future of open-source models in trajectory data mining. (3) As shown in Figure~\ref{fig:llm_comparison_porto}(c), o4-mini achieves the best results on both tasks while generating the shortest output responses. Its superior performance and efficiency can be attributed to its longest inference tokens. Given its dominant performance, we select o4-mini as the default MLLM backbone to report our performances. We show the output differences of different MLLMs in Appendix~\ref{append:interpretable}

\subsection{Efficiency of \our (RQ3)}
To answer \textbf{RQ3}, we evaluate the efficiency of \our by measuring inference latency, throughput, average time, and cost on different MLLM backbones. The throughput and average time are calculated based on the maximum allowed for the API provider's highest access tier (e.g., o4-mini has a throughput of 150M tokens per minute (TPM)). The results are summarized in Table~\ref{tab:efficiency}. 

As shown in the table, we have the following observations: (1) o4-mini is the most efficient, achieving the lowest latency and the highest throughput. In contrast, gemini-2.5-pro is the slowest model due to its significantly longer response time. (2) In terms of cost, gemma-3-14b is the most economical choice, which can be attributed to its smaller parameter size. The gemini-2.5-pro is the most expensive model owing to its large number of output tokens. (3) With the large scale deployment of MLLMs, the throughput increases and average processing time of each trajectory drops sharply, \ie, 0.030s per trajectory. These results highlight the trade-offs between inference speed, throughput, and cost to use MLLMs.

\begin{table}[tbp]
\centering
\renewcommand{\arraystretch}{0.8}
\setlength{\tabcolsep}{4.8pt}
\small
\caption{Efficiency of Different MLLMs.}
\vspace{-2ex}
\label{tab:efficiency}
\begin{tabular}{|l|c|c|c|c|}
\hline
Model & Latency & Throughput & AvgTime & Cost \\
\hline
gemma-3-14b     & 14.1  & 178.6 & 0.079 & \$0.002 \\
gemma-3-27b     & 11.2  & 268.1 & 0.042 & \$0.004 \\
qwen2.5-vl-32b  & 15.2  & 189.9 & 0.080 & \$0.006 \\
qwen-vl-max     & 10.4  & 277.3 & 0.038 & \$0.017 \\
qwen2.5-vl-72b  & 19.3  & 149.2 & 0.129 & \$0.018 \\
gemini-2.5-pro  & 32.0  & 89.9  & 0.356 & \$0.024 \\
o4-mini         & 9.23  & 312.5 & 0.030 & \$0.014 \\
\hline
\end{tabular}
\vspace{-3ex}
\end{table}

\begin{table}[h]
\centering
\caption{Ablation Study Results on the Chengdu Dataset.}
\label{tab:ablate_chengdu}
\vspace{-2ex}
{
\small
\renewcommand{\arraystretch}{0.88}
\setlength{\tabcolsep}{5pt} 
\begin{tabular}{|l|ccc|cc|}
\hline
\multirow{3}{*}{\textbf{Method}} & \multicolumn{5}{c|}{\textbf{Chengdu}} \\ \cline{2-6}
& \multicolumn{3}{c|}{\textbf{Travel Time Estimation}} & \multicolumn{2}{c|}{\textbf{Anomaly Detection}} \\ \cline{2-6}
& MAE $\downarrow$ & RMSE $\downarrow$ & MAPE $\downarrow$ & F1-score $\uparrow$ & PR-AUC $\uparrow$ \\ \hline

w/o image & 350.15 & 392.33 & 35.97 & - & - \\
w/o seg & 243.26 & 357.02 & 28.44 & 0.788 & 0.821 \\
w/o order & 330.23 & 368.12 & 32.87 & 0.952 & 0.953 \\
w/o POI & 194.06 & 247.79 & 19.92 & 0.940 & 0.942 \\
w/o road & 234.91 & 304.88 & 24.75 & 0.862 & 0.871 \\ \hline
\our & \textbf{133.43} & \textbf{154.67} & \textbf{16.34} & \textbf{0.980} & \textbf{0.980} \\ \hline
\end{tabular}
\vspace{-3ex}
}
\end{table}
\subsection{Ablation Study (RQ4)}\label{appen:ablation}
We compare \our with five ablations on two downstream tasks to analyze the effectiveness of the proposed components. We obtain \textbf{w/o image}, \textbf{w/o segment}, \textbf{w/o order}, \textbf{w/o POI}, and \textbf{w/o road}, by removing the visual modality, replacing the principled segmentation with a naive strategy, breaking the correspondence between textual descriptions and their visual components, removing the POI-centric view, and removing the road-network-centric view, respectively. The results are summarized in Table~\ref{tab:ablate_chengdu}. 

Removing the image modality (\textbf{w/o image}) causes the most significant performance drop on both tasks. This indicates that constructing visual representations is crucial for MLLMs to understand trajectory details. Breaking the order of visual-text pairs (\textbf{w/o order}), which removes temporal correspondence, also leads to a notable performance decrease. This means that temporal order is crucial to capture trajectory dynamics accurately. Replacing our semantic segmentation approach with a naive segmentation strategy (\textbf{w/o segment}) also leads to large performance degradation, demonstrating that the local detail view obtained through semantically coherent segmentation can provide crucial information. Furthermore, the lack of POI-centric view and road network-centric view also significantly reduces the performance. In a word, these ablation results support that the designed components of \our have positive contributions to the performance of \our.

\subsection{Reasoning Ability Analysis (RQ5)}
To answer \textbf{RQ5}, we analyze the reasoning ability of \our. Due to the space limit, we show the case figures and more detailed analysis in Appendix~\ref{append:reasoningability}. \our demonstrates strong reasoning abilities in two key aspects: (1) For \textbf{interpretable result}, \our provides detailed, step-by-step reasoning. For example, in the anomaly detection task, it not only identifies an anomalous trajectory but also precisely locates the anomalous segments and explains its conclusion by referencing the multimodal information. (2) For \textbf{multi-round reasoning}, the model can adapt its predictions based on real-time contextual updates. For instance, in the TTE task, after being notified of traffic congestion on a specific road, \our incorporates this new information to revise its initial travel time estimation. These capabilities show that \our functions as an interactive reasoning engine, not just a static prediction model.

\subsection{Sensitivity Analysis of $\theta$ (RQ6)}
\begin{figure}[tbp]
	\centering
\includegraphics[width=0.47\textwidth]{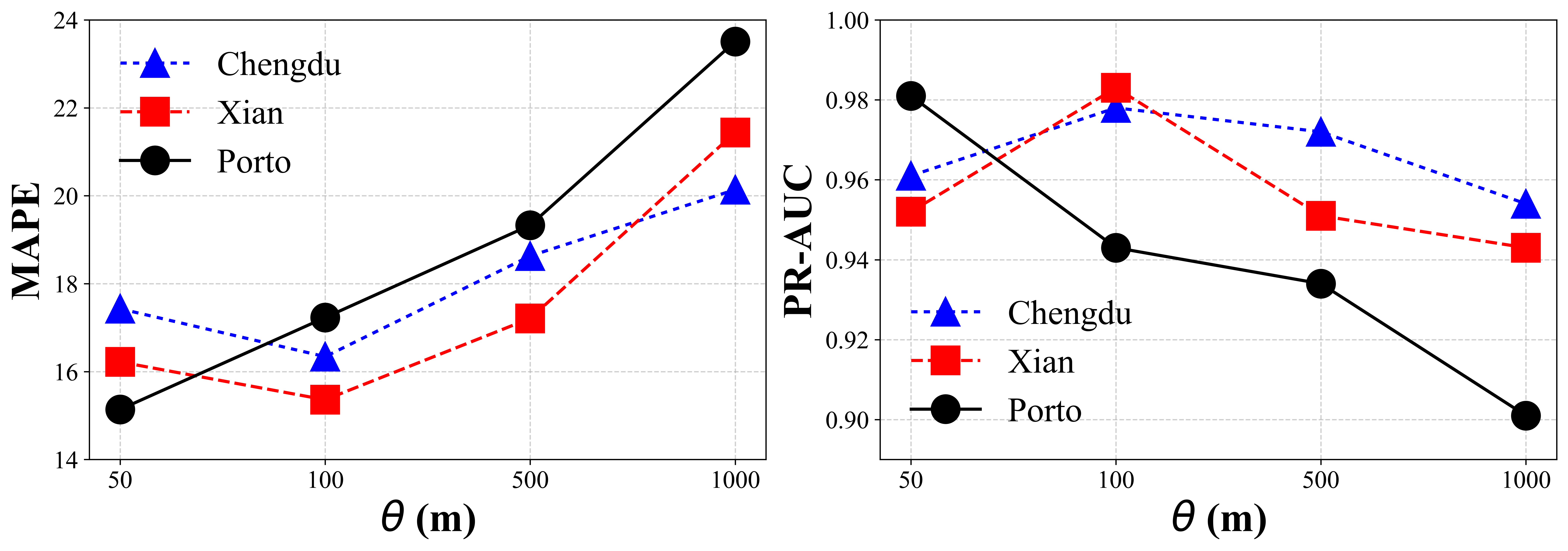}
    \vspace{-2ex}
\caption{Sensitivity Analysis.}	
\vspace{-4ex}
\label{fig:sensitivity}
\end{figure}

We conduct a sensitivity analysis on $\theta$ to investigate how different distance thresholds between context elements and the trajectory affect downstream task performance. According to the results presented in Figure~\ref{fig:sensitivity}, we have two key observations: (1) For all datasets, task performance consistently drops when the distance threshold exceeds 100m, indicating that larger thresholds introduce excessive irrelevant contextual information, which negatively impacts downstream tasks. (2) The optimal threshold differs by dataset: for Porto, the best results are achieved at 50m, while for Chengdu and Xian, 100m yields the highest performance. This difference reflects the variations in urban layouts—Chengdu and Xian have more regular road networks, making 100m a balanced choice between informativeness and noise. In contrast, Porto’s dense and irregular road structure favors a smaller 50m threshold to avoid interference from unrelated elements. More details are in Appendix~\ref{append:hyperparameters}.

\section{Conclusion}
In this paper, we propose a novel training-free framework \our which is the first work employing the reasoning ability of MLLMs for trajectory data mining. By effectively bridging the semantic gap of raw trajectory, \our integrates multiview contexts and transforms trajectories into multimodal sequences, enabling the use of the reasoning ability of MLLMs directly. Experiments on multiple representative tasks and cross-city datasets show that \our achieves performance improvements while maintaining strong generalization ability without any training or fine-tuning. We also release all the data responded by MLLMs, which can be used as the foundation for future research in training MLLMs specialized for trajectories and analyzing other spatial-temporal data.
\bibliographystyle{ACM-Reference-Format}
\bibliography{references}
\clearpage
\appendix
\section{Appendix}

\subsection{Details of Map-anchored Factors}\label{append:factor}

We define the three components used in the cost function $\text{Cost}(a,b)$ in detail as follows:

\begin{itemize}[leftmargin=*,itemsep=4pt]
    \item \textbf{Motion Homogeneity ($f_{speed}(a, b)$).} 
    This term measures the internal consistency of movement dynamics within the segment $T_{a..b}$. 
    Specifically, we compute the variance of instantaneous speeds between consecutive GPS points in the segment.
    A lower variance reflects smoother or more consistent motion, indicating that the segment corresponds to a coherent phase of movement (e.g., cruising or stationary).

    \item \textbf{Route Homogeneity ($f_{road}(a, b)$).}
    This term captures the structural consistency of the road types traversed in the segment. 
    We identify the sequence of road IDs that the trajectory crosses, and count the number of transitions between different roads.
    A smaller number of transitions indicates that the segment is located in a consistent road environment, which is consistent with the intuition that semantic stages should not cross heterogeneous road environments.

    \item \textbf{Segment Length Regularization ($f_{len}(a, b)$).}
  In order to avoid generating too short or simple segments, we introduce a regularization term based on segment length.
  Specifically, $f_{len}(a, b)$ is defined as the inverse of the number of points from $p_a$ to $p_b$, which encourages partitions to be long enough to be semantically meaningful while still maintaining internal consistency.

\end{itemize}

\subsection{Details of Semantic Descriptions}\label{append:description}
This section provides the detailed implementation of the semantic descriptions generated for each sub-trajectory $S_n$. As outlined in Equation~\ref{eq:text_token}, we extract key statistical and dynamic characteristics from the raw GPS points within a sub-trajectory. These features are then formatted into a structured textual template, which serves as a component of the multimodal input for the MLLM.

\paragraph{Time-based Features}
For each sub-trajectory $S_n$, which consists of a sequence of GPS points $\{p_1, \dots, p_k\}$, we extract fundamental temporal information. The \textbf{Start Time} is the timestamp of the first point ($t_1$), the \textbf{End Time} is the timestamp of the last point ($t_k$), and the \textbf{Duration} is the total time elapsed, calculated as ($t_k - t_1$) in seconds.

\paragraph{Distance-based Features}
The \textbf{Total Distance} of a sub-trajectory is calculated by summing the geographical distances between all consecutive GPS points. We use the Haversine formula to compute the great-circle distance between two points on Earth, ensuring an accurate measurement of the path traveled in meters.

\paragraph{Speed-based Features}
To capture the motion dynamics of the sub-trajectory, we compute a set of speed-related metrics. First, we calculate the instantaneous speed between each consecutive pair of points. Based on these values, we derive the following:
\begin{itemize}[leftmargin=4mm]
    \item \textbf{Average Speed:} Calculated as the Total Distance divided by the total Duration, providing a robust measure of the overall speed in m/s.
    \item \textbf{Maximum Speed:} The highest instantaneous speed recorded between any two consecutive points within the sub-trajectory.
    \item \textbf{Minimum Speed:} The lowest non-zero instantaneous speed recorded within the sub-trajectory.
\end{itemize}

\paragraph{Textual Template}
After computing the features described above, they are populated into a structured, human-readable text template. This format allows the MLLM to easily parse and comprehend the key characteristics of the trajectory segment. An example of the template is shown below:

\begin{verbatim}
--- Sub-trajectory Segment Description ---
Start Time: 2024-11-01 13:08:36
End Time: 2024-11-01 13:09:42
Duration (seconds): 66
Total Distance (meters): 410.8
Average Speed (m/s): 6.22
Maximum Speed (m/s): 8.51
Minimum Speed (m/s): 1.15
----------------------------------------
\end{verbatim}

\subsection{Experimental Settings}\label{append:settings}
This section provides a detailed experimental settings including datasets, task descriptions, baselines, and hyperparameter settings. 
\subsubsection{Dataset}\label{append:dataset}
To evaluate the cross-city generalization and cross-task universality of our framework, we conduct extensive experiments on four large-scale, real-world trajectory datasets, Xi'an, Chengdu, Porto and Geolife.

\begin{table}[!htbp]
\centering
\vspace{-2ex}
\caption{Dataset statistics.}
\vspace{-2ex}
\label{tab:dataset_summary}
\begin{tabular}{lcc}
\toprule
Dataset      & Trajectories & Time Span \\
\midrule
Chengdu   & 677,492            & 2016/11/01 -- 2016/11/30 \\
Xi'an     & 373,054            & 2016/11/01 -- 2016/11/30 \\
Porto     & 695,085           & 2013/07/01 -- 2014/07/01 \\
Geolife (Beijing) & 17,621 & 2007/04/01 -- 2012/08/01 \\
\bottomrule
\end{tabular}
\end{table}

\begin{itemize}[leftmargin=4mm]
    \item \textbf{Xi'an and Chengdu:} These two datasets contain taxi trajectories collected in Xi'an and Chengdu, China, during October and November 2016. They represent typical urban mobility patterns from ride-hailing services.
    
    \item \textbf{Porto:} This is a publicly available dataset of taxi trajectories from Porto, Portugal, originally released for a Kaggle competition\footnote{https://www.kaggle.com/c/pkdd-15-predict-taxi-service-trajectory-i}. Its widespread use in prior research makes it a standard benchmark for comparison.
    
    \item \textbf{GeoLife:} This dataset, collected by Microsoft Research Asia, contains GPS trajectories from 182 users over several years \cite{zheng2011geolife}. The dataset is highly diverse, comprising 17,621 trajectories with a total distance of over 1.2 million kilometers and a duration of more than 50,000 hours. It is ideal for evaluating generalization across different user behaviors and modes of transport.
\end{itemize}

\subsubsection{Tasks Description}\label{append:task}
We evaluate our framework on four representative downstream tasks that span both regression and classification settings:
\begin{itemize}[leftmargin=4mm]
    \item \textbf{Travel Time Estimation (TTE):} A regression task to predict the total travel time of a given trajectory. We use Mean Absolute Error (MAE), Root Mean Squared Error (RMSE), and Mean Absolute Percentage Error (MAPE) as evaluation metrics.
    \item \textbf{Anomaly Detection (AD):} A binary classification task to identify trajectories that deviate from normal mobility patterns. Following previous work~\cite{liu2020online}, we construct the anomaly dataset and report Accuracy (ACC), Precision (Prec), and the Area Under the Precision-Recall Curve (PR-AUC).
    \item \textbf{Mobility Prediction (MP):} A classification task that predicts the destination region of a trajectory. We use Top-1 Accuracy (ACC@1) and Top-5 Accuracy (ACC@5) for evaluation.
    \item \textbf{Transportation Mode Identification (TMI):} A multi-class classification task on the GeoLife dataset, aiming to identify the mode of transportation. We report Accuracy, Macro-F1, and Weighted-F1.
\end{itemize}

\noindent\textbf{Ground Truth.} Given the scarcity of real-world labeled data for trajectory anomaly detection, we follow previous works\cite{liu2020online,wang2024multi,shao2025fo} and adopt a disturbance-based approach to generate a synthetic anomaly dataset. This involves programmatically injecting two distinct types of anomalies—\textbf{detour} and \textbf{switch}—into normal trajectories.
\begin{itemize}[leftmargin=4mm]
    \item A \textbf{detour} is defined by two parameters: an anomaly probability $\alpha$ and a disturbance distance $d$. For instance, setting $\alpha = 0.1$ and $d = 3$ signifies that 10\% of the trajectory's points are displaced by a distance of 3 grid units from their original path.
    \item \textbf{switch} simulates a more drastic deviation, controlled by the trajectory split point $\mu$. An instance with $\mu = 0.3$ indicates that a trajectory follows its original route $r_1$ for the first 30\% of its course, before abruptly switching to an entirely different route $r_2$ for the remainder of the journey.
\end{itemize}
In our experiments, the overall proportion of anomaly injection is consistently set to 5\%.

\subsubsection{Baselines}\label{append:baselines}
For each task, we choose three kinds of baselines, \ie, task-specified models, trajectory foundation models and LLM-based models. 

\noindent\textbf{Baselines for TTE and MP}
The baselines for Travel Time Estimation (TTE) and Mobility Prediction (MP) include the following methods:
\begin{itemize}[leftmargin=*,itemsep=2pt]
    \item \textbf{Traj2Vec~\cite{yao2017trajectory}:} An RNN-based seq2seq model that converts a GPS trajectory into a feature sequence to learn representations, originally for the trajectory clustering task.

    \item \textbf{T2Vec~\cite{li2018deep}:} A denoising auto-encoder that pre-trains a model by reconstructing original trajectories from their low-sampling-rate counterparts.

    \item \textbf{Trember~\cite{fu2020trembr}:} An RNN-based seq2seq model designed to learn trajectory representations by recovering the road segments and timestamps of the input trajectories.

    \item \textbf{CTLE~\cite{lin2021pre}:} A method that pre-trains a bi-directional Transformer with two Masked Language Model (MLM) tasks focusing on location and hour predictions.

    \item \textbf{Toast~\cite{chen2021robust}:} This framework utilizes a context-aware node2vec model to generate road segment representations and trains the model with an MLM-based task and a sequence discrimination task.

    \item \textbf{TrajCL~\cite{chang2023contrastive}:} A contrastive learning framework that introduces a dual-feature self-attention-based encoder and trains the model in a contrastive style using the InfoNCE loss.

    \item \textbf{MMTEC~\cite{lin2023pre}:} A multi-modal method that learns trajectory representations using an attention-based discrete encoder and a NeuralCDE continuous encoder to extract travel behavior and continuous spatial-temporal correlations.

    \item \textbf{JGRM~\cite{ma2024more}:} A multi-modal method that learns trajectory representations from the joint perspectives of free-space (GPS) and road networks, using a combination of contrastive loss, MLM loss, and alignment loss.

    \item \textbf{START~\cite{jiang2023self}:} A self-supervised framework that includes a time-aware trajectory encoder and a Graph Attention Network (GAT) and is trained with both an MLM task and a SimCLR-based contrastive task.
    
    \item \textbf{UniTR~\cite{zhao2025unitr}:} A unified framework for joint representation learning of road networks and trajectories, which uses a hierarchical propagation mechanism to model their many-to-many interactions and a triple-level contrastive loss for optimization.

    \item \textbf{BLUE~\cite{zhou2025blue}:} An encoder-decoder model with a pyramid structure that gradually reduces GPS coordinate precision to create hierarchical patches, thereby capturing both fine-grained details and high-level travel patterns.

    \item \textbf{PLM4Traj~\cite{zhou2024plm4traj}:} A model that leverages Pre-trained Language Models (PLMs) for trajectory analysis by incorporating a novel trajectory prompt and a semantic embedder to understand movement patterns and travel purposes.

    \item \textbf{Path-LLM~\cite{wei2025path}:} A multi-modal path representation learning model that integrates Large Language Models (LLMs) to interpret both topological and textual path data, using contrastive pretraining for alignment and a dynamic fusion module.

    \item \textbf{MM-Path~\cite{xu2025mm}:} A multi-modal, multi-granularity framework designed to learn a generic path representation by systematically aligning and fusing information from both road network-based paths and image-based (e.g., remote sensing) paths.
\end{itemize}

\noindent\textbf{Baselines for AD}
For Anomaly Detection (AD), we compare against the following methods:
\begin{itemize}[leftmargin=*,itemsep=2pt]
    \item \textbf{iBAT~\cite{zhang2011ibat}:} An isolation-based method that detects anomalous trajectories by evaluating how much a target trajectory can be isolated from a set of reference trajectories.

    \item \textbf{GM-VSAE~\cite{liu2020online}:} A Gaussian mixture variational sequence autoencoder, which is an RNN-based sequence-to-sequence model that handles complex trajectory distributions by employing a Gaussian mixture model in the latent space.

    \item \textbf{ATROM~\cite{gao2023open}:} A probabilistic metric learning model that focuses on open anomalous trajectory recognition, aiming to identify which type of anomaly a given trajectory exhibits.

    \item \textbf{DeepTEA~\cite{han2022deeptea}:} A state-of-the-art time-dependent method that combines dynamic traffic conditions with trajectory patterns to effectively and efficiently detect online trajectory outliers.

    \item \textbf{MST-OATD~\cite{wang2024multi}:} A multi-scale model designed for online anomalous trajectory detection that considers both spatial and temporal aspects of trajectories to extract features at multiple scales.
    
    \item \textbf{FOTraj~\cite{shao2025fo}:} An LLM-driven framework for fine-grained and noise-resilient trajectory anomaly detection, which converts trajectories into spatial-temporal graphs and uses a Patch-Gated Robustification (PGR) module to enhance noise resilience.
\end{itemize}

\noindent\textbf{Baselines for TMI}
For Transportation Mode Identification (TMI), the baselines are as follows:
\begin{itemize}[leftmargin=*,itemsep=2pt]
    \item \textbf{SECA~\cite{dabiri2019semi}:} A deep semi-supervised convolutional autoencoder (SECA) architecture that automatically extracts features from GPS segments and leverages both labeled and unlabeled data for classification.

    \item \textbf{TrajODE~\cite{liang2021modeling}:} A state-of-the-art method that couples the continuity of Neural Ordinary Differential Equations (ODE) with the robustness of latent variables to model trajectories.

    \item \textbf{TrajFormer~\cite{liang2022trajformer}:} A transformer-based architecture that generates continuous point embeddings to handle irregular spatial-temporal intervals and uses a squeeze function to improve efficiency for long trajectories.

    \item \textbf{RED~\cite{zhou2024red}:} A self-supervised Transformer-based framework that employs a road-aware masking strategy, a spatial-temporal-user joint embedding scheme, and a dual-objective (next segment prediction and full trajectory reconstruction) to learn comprehensive representations.
\end{itemize}

\subsection{Hyperparameter Analysis}\label{append:hyperparameters}
In this section, we analyze the sensitivity of our framework to its key hyperparameter, the context filtering distance threshold $\theta$, which is defined in Equation~\ref{eq:poi_filter}. This parameter is crucial as it controls the amount of contextual information (e.g., POIs, traffic lights) included in the visual representations fed to the MLLM. 

\begin{figure}[htbp]
  \centering  \includegraphics[width=0.43\textwidth]
  {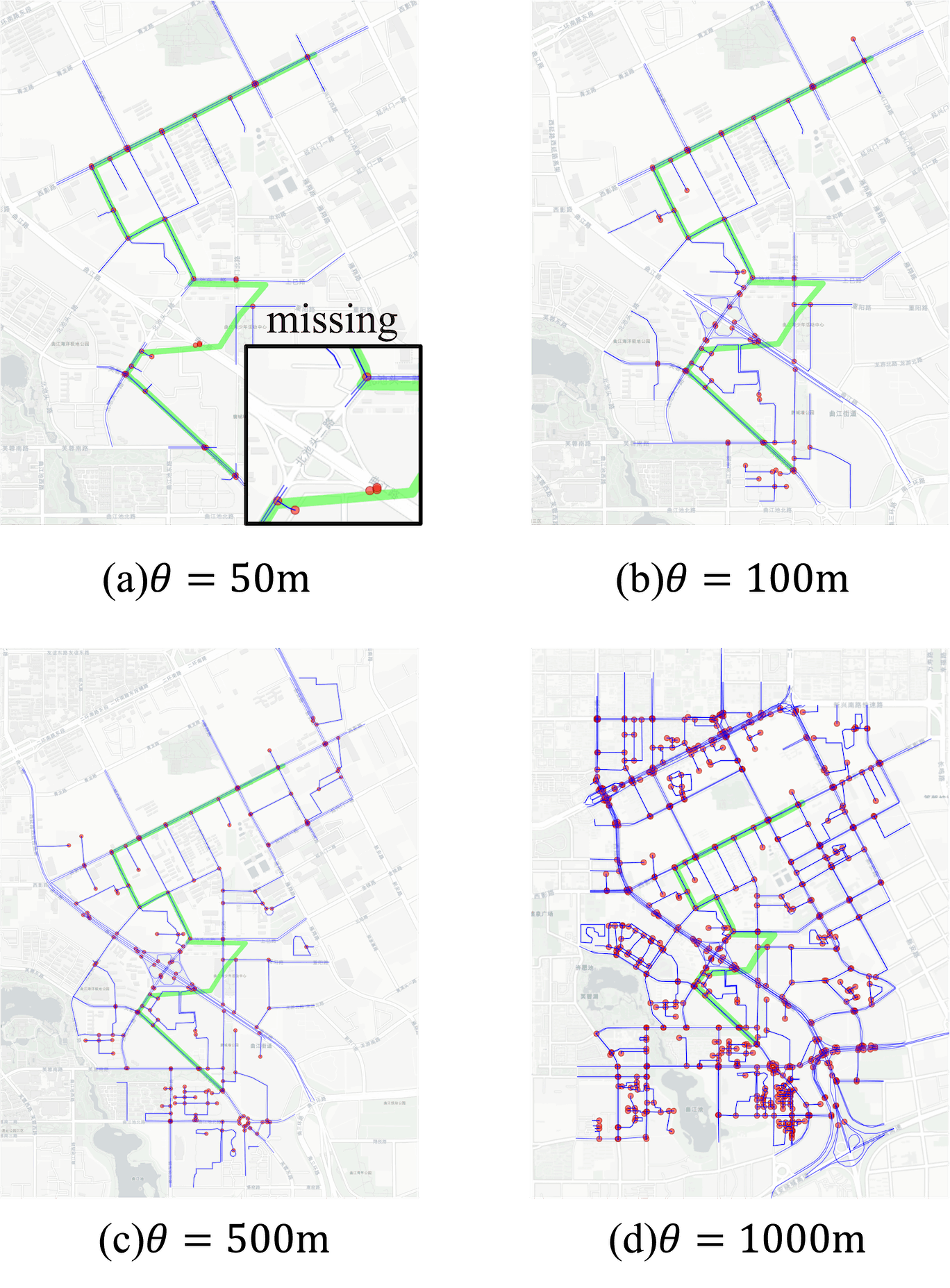}
    \vspace{-2ex} 
  \caption{Visual comparison of abnormal trajectories under different distance thresholds.\label{fig:hyper_example}}
  \vspace{-2ex} 
\end{figure}

The choice of $\theta$ involves a critical trade-off. A threshold that is too small may exclude relevant nearby elements that are important for reasoning. For example, in an anomaly detection task as illustrated in Figure~\ref{fig:hyper_example}(a), a restrictive threshold could fail to visualize an existing connecting road. This visual omission might mislead the MLLM to incorrectly perceive the path as disconnected and thus flag a normal trajectory as an anomaly. Conversely, a threshold that is too large can introduce an excessive number of irrelevant elements into the map view. This creates visual clutter and acts as noise, potentially overwhelming the MLLM and degrading its performance.

To investigate this impact empirically, we conducted a sensitivity analysis by varying the value of $\theta$. We evaluated the performance on the Travel Time Estimation (TTE) task, measuring the Mean Absolute Error (MAE). The results of this analysis are illustrated in Figure~\ref{fig:hyper_example}(b).

As shown in Figure~\ref{fig:hyper_example}(b), the model's performance (lower MAE is better) initially improves as the distance threshold increases from a very restrictive value (e.g., 50m). This is because a larger radius allows the model to incorporate more relevant contextual cues. However, after reaching an optimal point (e.g., 100m for Chengdu and Xian), the performance begins to decline steadily. This trend confirms our hypothesis: overly large values for $\theta$ lead to the inclusion of too much irrelevant information, which introduces noise and ultimately hinders the MLLM's reasoning process.

Based on this analysis, we set $\theta$ tailored to the scale of each city's road network. For the Xian, Chengdu, and Geolife datasets, which feature similarly larger road networks, we selected $\theta_{\text{distance}} = 100$ meters. In contrast, for the Porto dataset with its smaller transportation network, we chose a more constrained value of $\theta_{\text{distance}} = 50$ meters. This approach ensures that each setting strikes an effective balance between providing sufficient contextual information while minimizing distracting extraneous elements.

\subsection{Reasoning Ability of MLLM's Outputs}\label{append:reasoningability}

\subsubsection{Multi-round reasoning}\label{append:reasoning}
\begin{figure*}[t]
    \centering
    \includegraphics[width=0.82\textwidth]{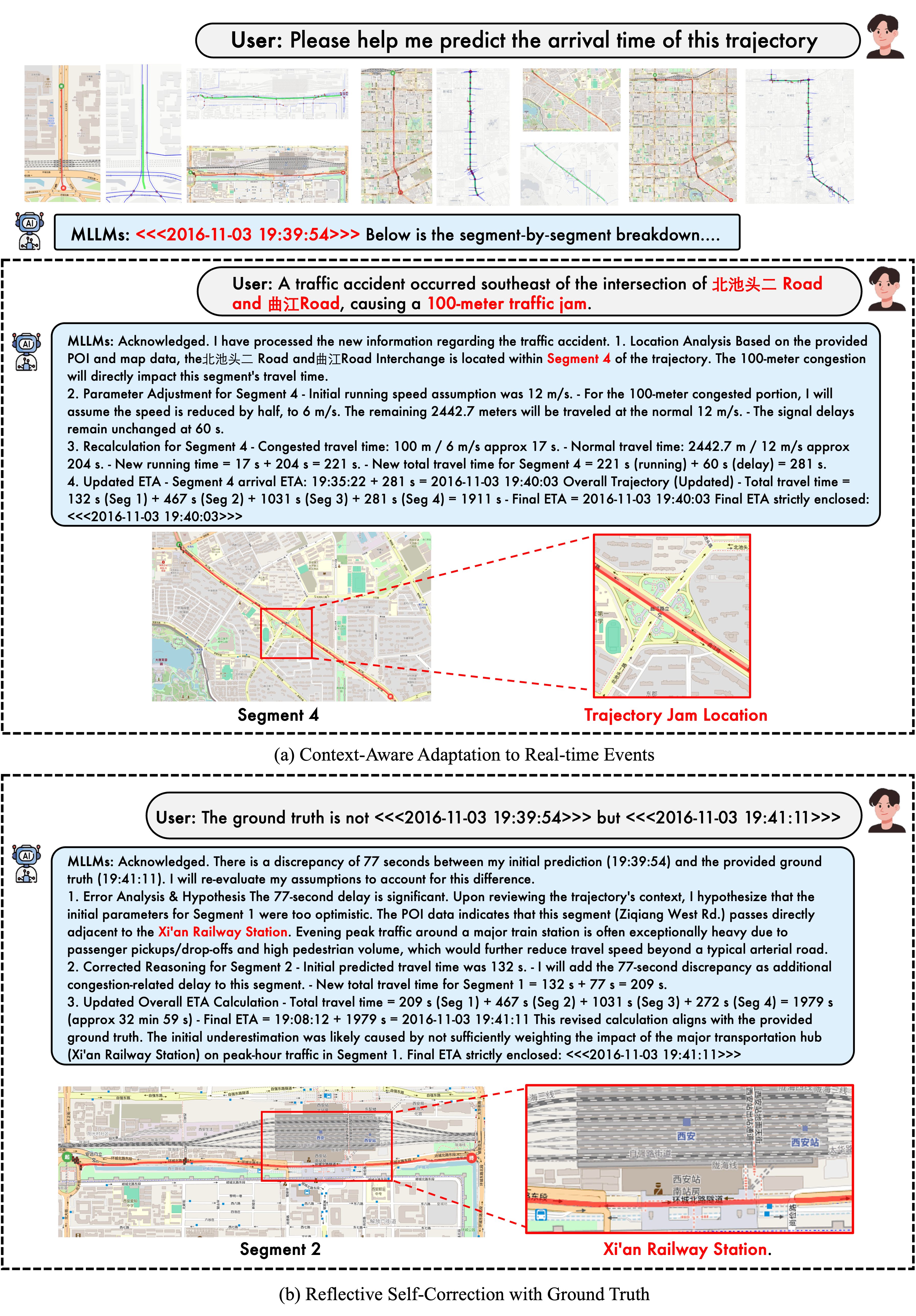}
    \caption{The result of \our.}
    \label{fig:casestudy}
\end{figure*}

To further assess the advanced capabilities of \our, we demonstrate its capacity for dynamic, multi-round reasoning. Unlike static models that produce a one-off output, our MLLM-based framework can engage in interactive sessions, adapting its analysis based on new information. We present two case studies drawn from the Travel Time Estimation task, as illustrated in Figure~\ref{fig:casestudy}. These cases highlight two key abilities: (1) adapting predictions in response to real-time contextual updates, and (2) performing reflective self-correction when presented with ground-truth feedback.

\paragraph{Context-Aware Adaptation to Real-time Events}
The first case study demonstrates the model's ability to adjust its reasoning when provided with new, real-time information. The model initially provides a complete ETA prediction based on the multimodal trajectory data. We then introduce a new piece of information—a traffic accident—to test if the model can correctly locate the event, update its parameters, and recalculate the ETA.

\paragraph{Reflective Self-Correction with Ground Truth}
The second case study showcases the model's ability to reflect on its own prediction error. After the initial prediction is made, we provide the ground-truth arrival time and ask the model to explain the discrepancy. This mimics a feedback loop where the model can learn and refine its reasoning based on outcomes.

\subsubsection{Interpretable Results}\label{append:interpretable}
A key advantage of our framework is its ability to generate human-readable, interpretable results, leveraging the inherent reasoning capabilities of the MLLM backbone. This goes beyond simply outputting a prediction; the model can articulate the step-by-step logic behind its conclusions, making its outputs transparent and trustworthy. To demonstrate this, we present case studies for our primary downstream tasks, showcasing two core capabilities: sophisticated instruction following and expert-level logical reasoning.

\begin{figure*}[htbp]
  \centering  \includegraphics[width=0.99\textwidth]{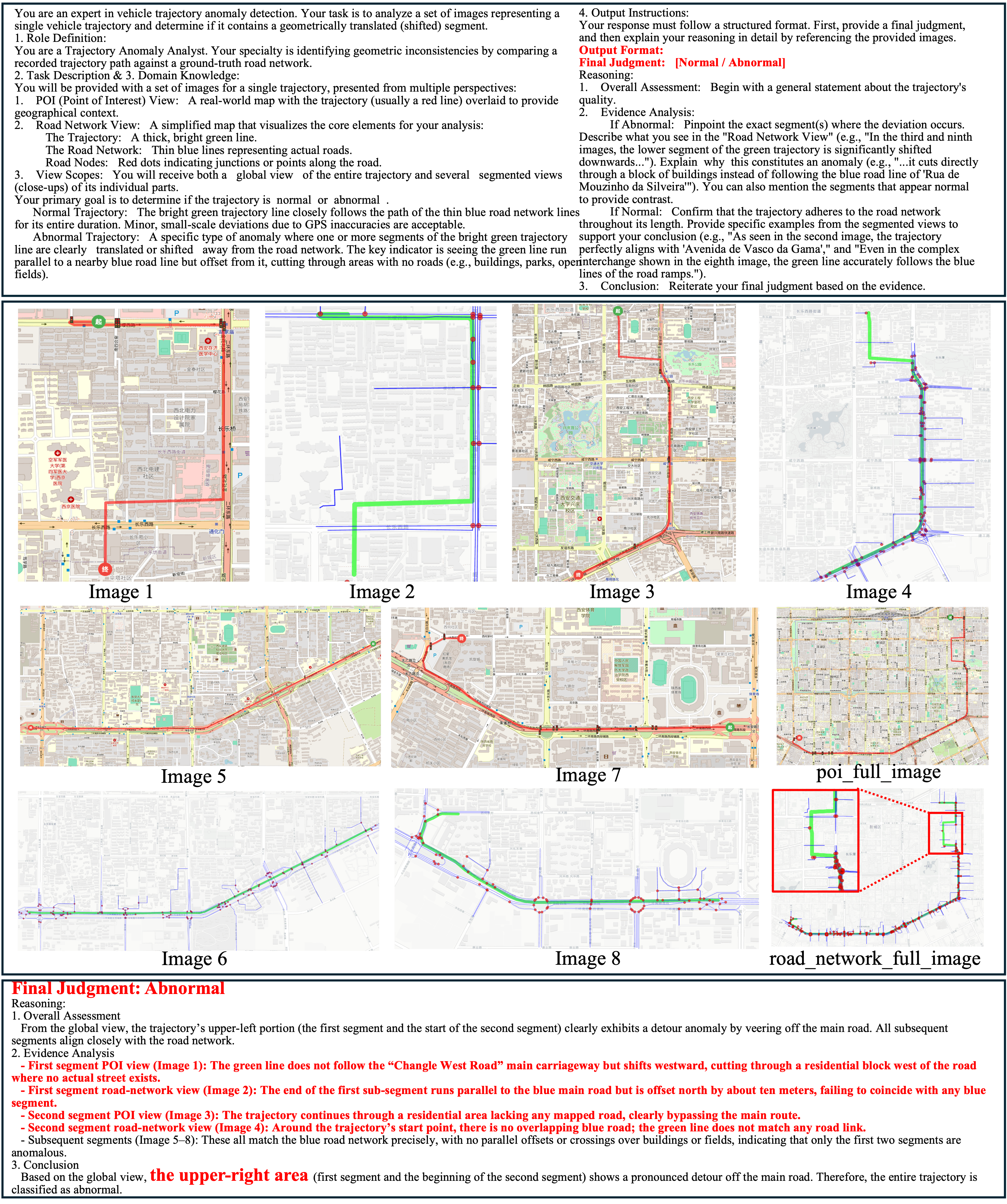}
  \caption{The result of Anomaly Detection. Ground Truth: Anomaly\label{fig:append_anomaly}}
\end{figure*}

\begin{figure*}[htbp]
  \centering
  \includegraphics[width=0.98\textwidth]{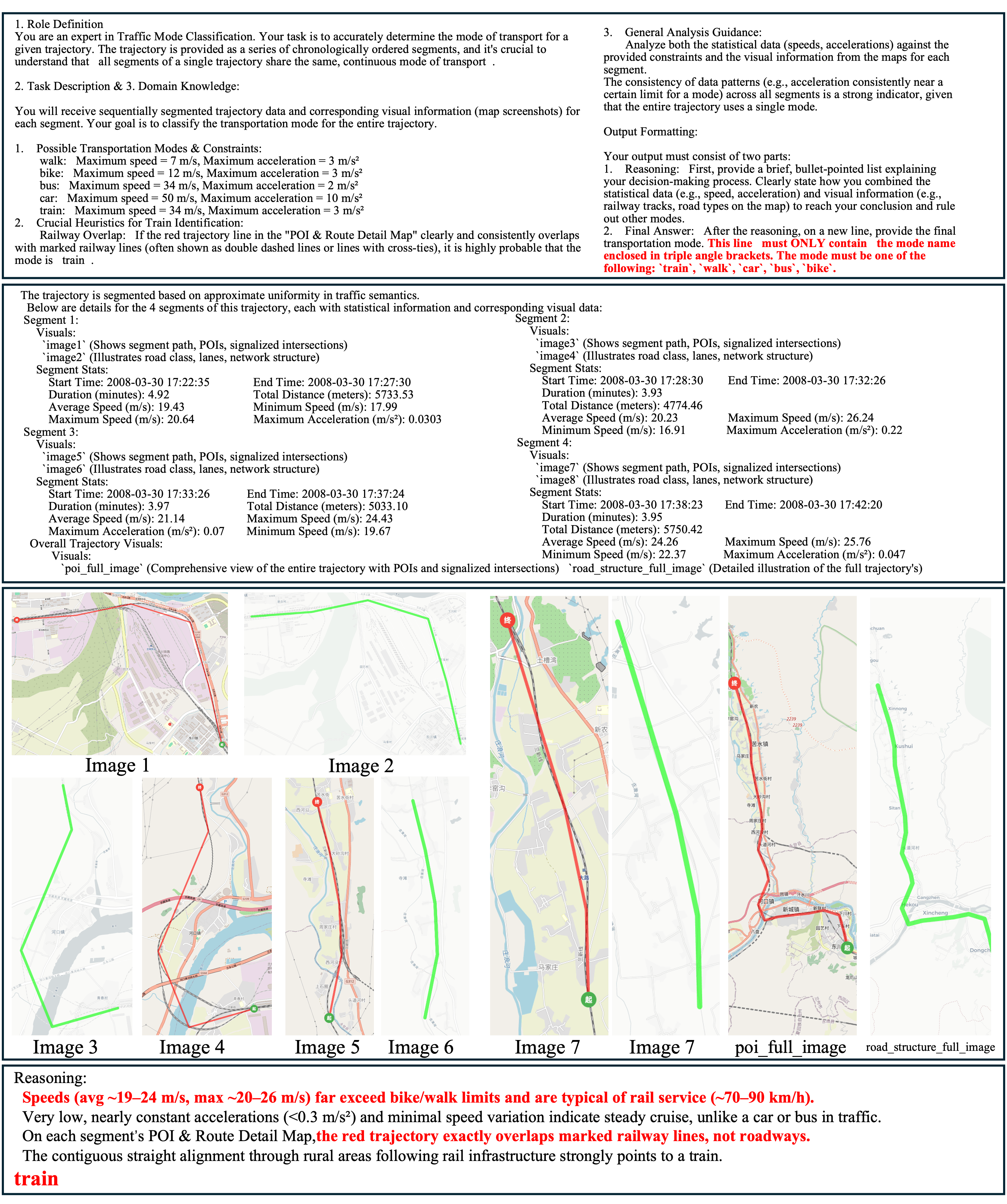}
  \caption{The result of  Transportation Mode Identification. Ground Truth: Train\label{fig:append_tmi}}
\end{figure*}

\paragraph{Sophisticated Instruction Following}
The MLLM backbone exhibits a remarkable ability to adhere to complex and highly structured instructions defined in the system prompt. This ensures that the model's output is not only accurate but also consistently formatted, making it reliable for automated pipelines.
\begin{itemize}[leftmargin=4mm]
    \item \textbf{Case 1: Structured Formatting.} In the Transportation Mode Identification task (Figure~\ref{fig:append_tmi}), the system prompt provides strict formatting rules for the ``Final Answer''. It dictates that the answer must appear on a new line after the reasoning and contain \textit{only} the mode name from a predefined list (e.g., `train`, `walk`, etc.). The output from o4-mini flawlessly adheres to these rules, showcasing its strong instruction-following capabilities.
    \item \textbf{Case 2: Complex Hierarchical Instructions.} The Anomaly Detection task (Figure~\ref{fig:append_anomaly}) requires an even more complex, hierarchical output as defined in its system prompt: a ``Final Judgment'' followed by a three-part ``Reasoning'' section (Overall Assessment, Evidence Analysis, and Conclusion). As the figure shows, the o4-mini model meticulously follows this multi-level structure, generating a well-organized analytical report that is easy for a human expert to parse. This showcases a high degree of instruction-following capability on complex, nested requirements.
\end{itemize}

\begin{figure*}[htbp]
  \centering
  \includegraphics[width=0.99\textwidth]{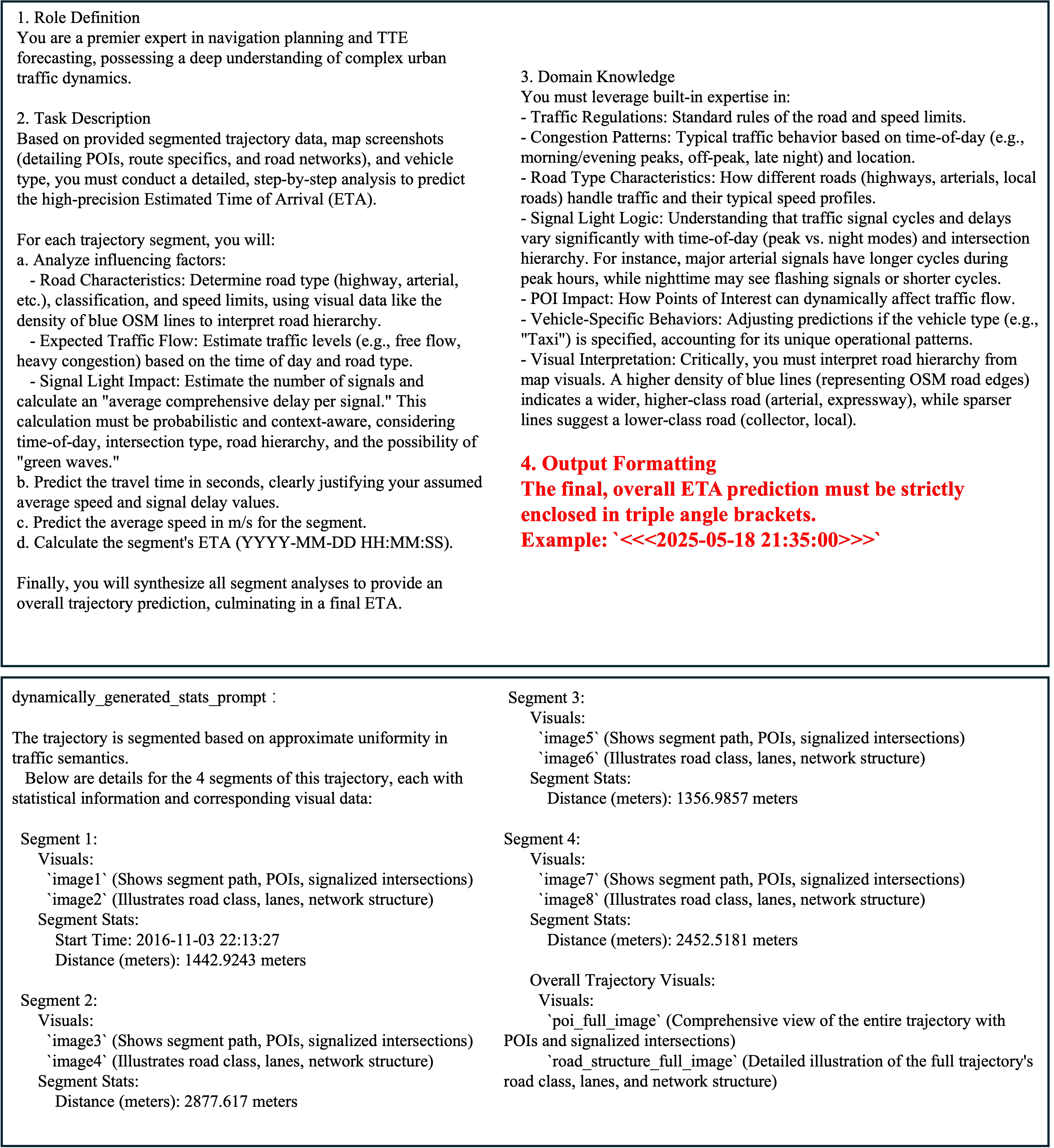}
  \caption{The system prompt of Travel Time Estimation.\label{fig:sysTTE}}
  \vspace{1cm} 
\end{figure*}

\begin{figure*}[htbp]
  \centering
  \includegraphics[width=0.9\textwidth]{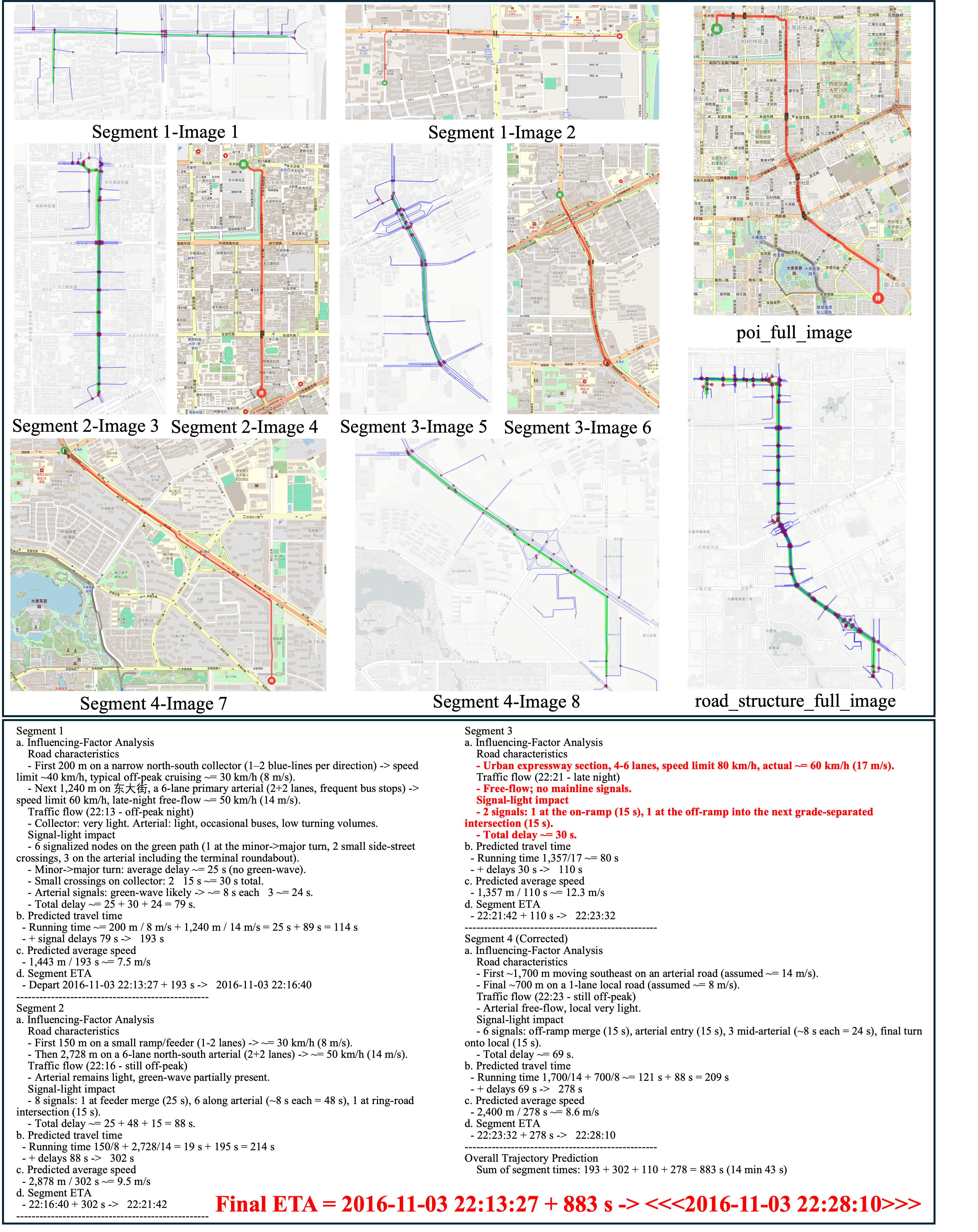}
  \caption{The result of Travel Time Estimation. Ground Truth: 2016-11-03 22:29:01\label{fig:tte}}
\end{figure*}

\paragraph{Correctness and Logical Reasoning}
More importantly, the model does not just follow formats blindly; it produces correct answers by executing a logical, step-by-step analysis that mimics human expert reasoning. It effectively synthesizes multimodal information to draw sound conclusions.
\begin{itemize}[leftmargin=4mm]
    \item \textbf{Case 1: Multimodal Synthesis.} In the Transportation Mode Identification task (Figure~\ref{fig:append_tmi}), the model correctly identifies the mode as ``train.'' Its reasoning demonstrates a powerful synthesis of information: it first uses statistical data (``Speeds (avg~19-24 m/s, max ~20-26 m/s) far exceed bike/walk limits'') to eliminate other modes, and then confirms its hypothesis using visual evidence from the user prompt's images (``the red trajectory exactly overlaps marked railway lines''). This process of elimination and confirmation is a hallmark of robust analytical reasoning.

    \item \textbf{Case 2: Complex Step-by-Step Calculation.} For Travel Time Estimation, the system prompt (Figure~\ref{fig:sysTTE}) instructs the model to act as a navigation expert and perform a detailed, segment-by-segment analysis. The model's output (Figure~\ref{fig:tte}) shows it executing this complex logic flawlessly. It breaks the journey into four segments, analyzes road characteristics (e.g., ``6-lane primary arterial''), traffic flow (``late-night free-flow''), and signal impacts (``2 signals: 1 at the on-ramp (15s)...''), and then performs explicit calculations (``Running time 1,357 / 17 ~= 80s'') before summing the results. This demonstrates that the model is performing genuine, step-by-step reasoning rather than simply pattern matching.
\end{itemize}

\begin{figure*}[t]
    \centering
    \includegraphics[width=0.99\textwidth]{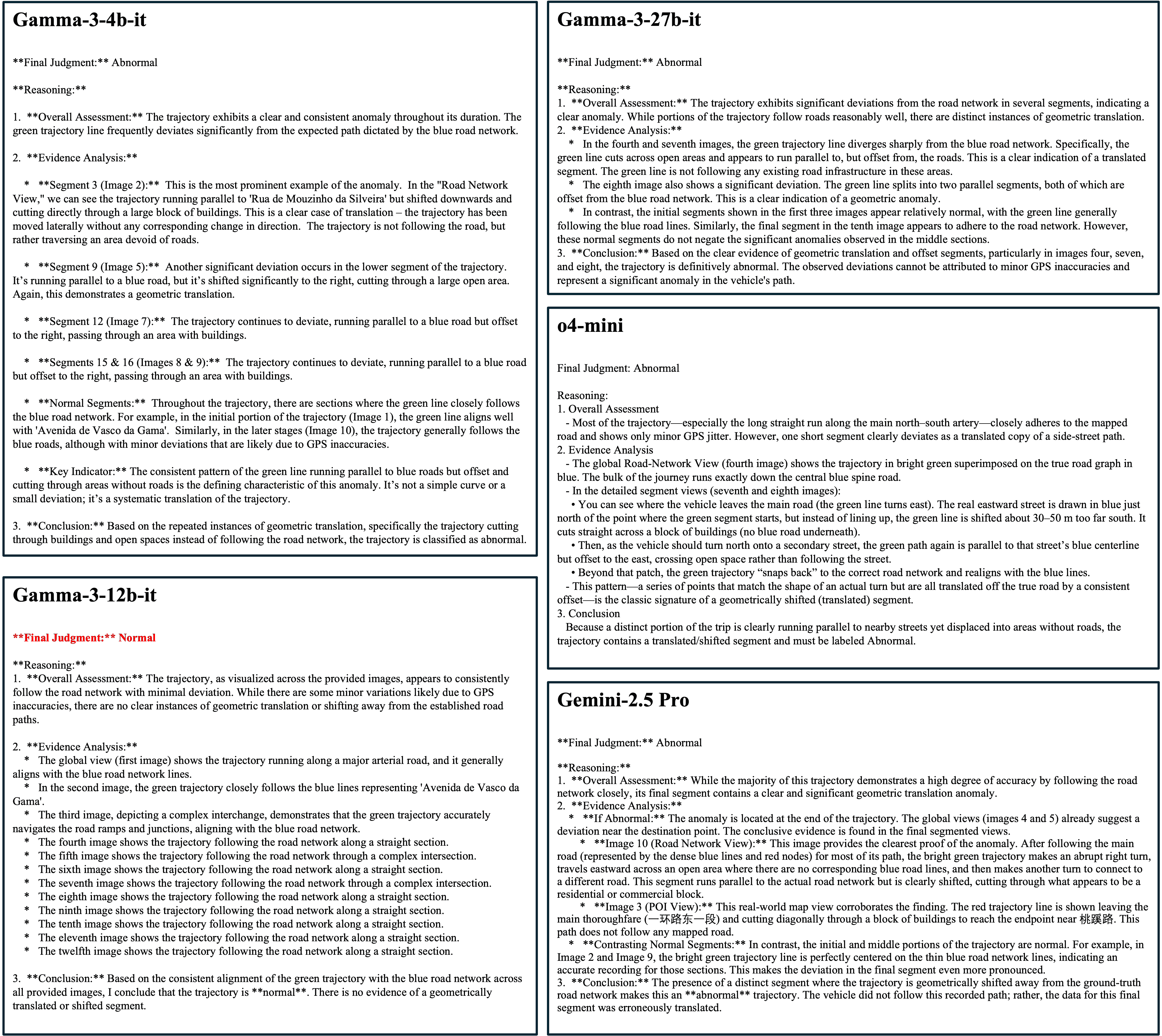}
    \caption{Qualitative comparison of outputs from different MLLM backbones for the Anomaly Detection task. The same input was provided to each model.}
    \label{fig:ad_model_comparison}
\end{figure*}

To further underscore the advanced reasoning quality of our chosen o4-mini backbone, we provide a qualitative comparison of outputs from different MLLMs for the same anomaly detection task in Figure~\ref{fig:ad_model_comparison}. This comparison reveals that not all models are equally adept at this specialized task. While most models like Gamma-3-27b-it and Gemini-2.5 Pro correctly identify the trajectory as abnormal, their reasoning, though valid, is less precise. For instance, the Gamma-3-12b-it model completely fails, misclassifying the trajectory as ``Normal.'' In stark contrast, o4-mini provides a diagnosis with expert-level precision, identifying the anomaly not just as a deviation, but as a ``translated copy of a side-street path'' and a ``classic signature of a geometrically shifted (translated) segment.'' This superior analytical depth validates our choice of backbone and highlights the effectiveness of our overall framework.


\clearpage
\end{document}